\documentclass{aa}
\usepackage{amsmath}
\usepackage{graphicx}
\usepackage{breqn}  
\usepackage[mathscr]{euscript} 
\usepackage{comment}
\usepackage[utf8]{inputenc}

\usepackage{booktabs}
\usepackage{adjustbox}
\usepackage{tablefootnote}

\usepackage{tablefootnote}
\usepackage{caption}
\usepackage{subcaption}
\usepackage{ulem}

\usepackage{placeins}

\usepackage{multicol}

\usepackage{multirow}

\usepackage[]{hyperref}
\hypersetup{unicode=true, colorlinks=true, linkcolor=[rgb]{0.53, 0.15, 0.34}, citecolor=[rgb]{0.0, 0.27, 0.42}, filecolor=[rgb]{1.0, 0.13, 0.32}, urlcolor=[rgb]{0.53, 0.15, 0.34}}

\usepackage{natbib} 
\bibpunct{(}{)}{;}{a}{}{,} 

\usepackage{txfonts}
\graphicspath{{Graphics/}}

\begin{document}

   \title{Broadband emission of microquasar remnants}
   \author{Leandro Abaroa\inst{1,2}\thanks{Corresponding author: \texttt{leandroabaroa@gmail.com}}, 
          Gustavo E. Romero\inst{1,2},
           \and Valentí Bosch-Ramon\inst{3}
           }

  \institute{Instituto Argentino de Radioastronom\'ia (CCT La Plata, CONICET; CICPBA; UNLP), C.C.5, (1894) Villa Elisa, Argentina \and Facultad de Ciencias Astron\'omicas y Geof\'{\i}sicas, Universidad Nacional de La Plata, B1900FWA La Plata, Argentina \and Departament de F\'isica Qu\`antica i Astrof\'isica, Institut de Ci\`encies del Cosmos (ICC), Universitat de Barcelona (IEEC-UB), Mart\'i i Franqu\`es 1, E08028 Barcelona, Spain}

   \date{Received / Accepted}

%-------------------------------------------------------

\abstract{Microquasar remnants (MQRs), the long-lived cocoons inflated by extinct microquasar jets, have recently been proposed as hidden Galactic PeVatrons capable of producing ultra-high-energy gamma rays without an active central engine. While hadronic interactions can account for bright gamma-ray emission from nearby clouds, the direct detection of MQRs remains challenging because their intrinsic emission is expected to be extended and of low surface brightness. In this work, we explore the broadband emission of MQRs by focusing on the leptonic component confined within the cocoon and on particle interactions in the shocked shell surrounding it. We model the injection and time-dependent transport of relativistic particles, including stochastic re-acceleration driven by internal turbulence, treated as a second-order Fermi process. We consider both sub-Eddington and super-Eddington microquasar systems and compute the resulting non-thermal emission from radio to gamma-ray energies, together with the thermal soft X-ray emission produced in the shocked shell. In the fiducial super-Eddington case, the intrinsic emission reaches peak values of \(\nu L_\nu \sim 10^{35}-10^{36}\,{\rm erg\,s^{-1}}\), whereas sub-Eddington remnants are typically several orders of magnitude fainter. At 1.3 GHz, the modeled cocoon surface brightness is of order \(\Sigma_\nu \sim 10^{-19}\,{\rm W\,m^{-2}\,Hz^{-1}\,sr^{-1}}\) for young powerful remnants and decreases rapidly as the remnant evolves. We find that the direct detectability of MQRs is therefore controlled mainly by surface brightness rather than by integrated luminosity. Powerful remnants may be detectable as extended synchrotron radio cocoons and shell-dominated soft X-ray structures under favorable conditions, whereas sub-Eddington remnants are expected to be much harder to identify directly. Our results suggest that MQRs may constitute a hidden population of extended Galactic non-thermal sources, whose identification requires multiwavelength searches combining morphology, surface brightness, and environmental context.}

\authorrunning{Abaroa et al.}
\titlerunning{Broadband emission of microquasar remnants}

\maketitle
\section{Introduction}\label{sec: intro}

The origin of cosmic rays (CRs) with energies around the knee of the spectrum remains one of the central problems of high-energy astrophysics. In recent years, the discovery of ultra-high-energy (UHE; $E_\gamma \gtrsim 100\,\mathrm{TeV}$) gamma-ray sources by the Large High Altitude Air Shower Observatory (LHAASO) has provided compelling evidence for the existence of PeVatrons in the Milky Way. While supernova remnants are long-standing candidates, recent observations \citep{LHAASO2024a,LHAASO2024b} and energetic arguments \citep{Wang2025ApJ} suggest that they may struggle to account for the most extreme energies. This motivates exploring alternative accelerators capable of sustaining PeV particle populations over extended timescales.

Accreting compact binaries with powerful outflows have emerged as promising contributors to the Galactic PeVatron population. In particular, recent theoretical and observational studies indicate that super-Eddington microquasars (MQs), characterized by sustained jet powers of $\sim10^{39}$--$10^{41}\,\mathrm{erg\,s^{-1}}$, can accelerate particles up to energies of several PeV, placing them among the most powerful Galactic accelerators \citep{Abaroa_etal2024(S26),Yue_etal_2024,Fujita_etal_2025,Peretti2025,Zhang_etal_2025, peretti2026particleaccelerationrecollimationshocks, li_2026, Vecchiotti_etal_2026, oliveranieto2026extremeparticleaccelerationxray}. This picture is further supported by the association of several LHAASO and the High-Altitude Water Cherenkov Observatory (HAWC) sources with MQs \citep{HAWK_SS433_2018Nature,V46412024Nature,LHAASO_2025}.

In a recent work, we proposed that the long-lived remnants of extinct MQs---hereafter microquasar remnants (MQRs)---can act as hidden PeVatrons \citep{Abaroa_etal_2026MQR}. In this scenario, relativistic particles injected during the active phase of the MQ remain confined within the overpressured cocoon long after accretion and jet production have ceased. Particles diffusing from these fossil CR reservoirs can illuminate dense gas structures through hadronic interactions \citep{2005A&A...432..609B}, producing bright UHE gamma-ray emission even in the absence of any currently active engine. Because the central binary becomes radiatively quiet, MQRs are expected to be difficult to identify through conventional multiwavelength signatures. Consequently, the irradiated clouds will likely appear as unidentified Galactic LHAASO sources.

The detection of MQRs poses a major observational challenge. As discussed in \cite{Abaroa_etal_2026MQR}, the radio surface brightness of MQR cocoons powered by primary electrons interacting with ambient fields is expected to be low, owing to the extended source size and the likely hadron-dominated non-thermal particle content. As a result, MQRs may evade detection in existing radio surveys, while their gamma-ray emission can appear spatially offset from the original accelerator if dominated by interactions with nearby clouds. Identifying robust observational diagnostics of MQRs therefore requires an exploration of their leptonic emission channels and their broadband, spatially resolved signatures. In particular, while our previous study focused on hadronic interactions as indirect tracers of MQRs, it did not address whether the remnant itself can be detected as an extended source through its intrinsic emission.

In this work, we extend the MQR framework of \cite{Abaroa_etal_2026MQR} by focusing on the cocoon itself as a diffuse, extended emitter, and on the injection, transport, and radiative losses of relativistic electrons confined within it. We investigate how primary electrons, injected at jet termination regions during the MQ phase, evolve under the combined effects of synchrotron losses in the cocoon magnetic field and inverse Compton scattering on ambient photon fields. In addition, we incorporate stochastic re-acceleration driven by internal turbulence, modeled as a second-order Fermi process —a mechanism neglected in our previous study. Such re-acceleration is expected to operate naturally in magnetized, turbulent cocoons and can modify the high-energy electron population and its radiative output (\citealp[e.g.][]{Romero&Muller&Roth2018Starburst}). This approach is complementary to cloud-illumination scenarios and aims to assess whether MQRs can be directly identified through their intrinsic emission. In our model, stochastic re-acceleration does not replace first-order acceleration at the jet termination shock, but acts as a secondary process that reshapes the electron distribution during the remnant phase.

We investigate both sub-Eddington and super-Eddington MQs. For each case, we calculate the time-dependent particle distributions and predict the resulting synchrotron and inverse Compton emission to assess the detectability of MQRs in radio and gamma rays. In addition, we estimate the thermal soft X-ray emission produced in the shocked shell of the cocoon as it expands into the interstellar medium (ISM). We also estimate the non-thermal emission arising from hadronic and leptonic interactions with the shell. Furthermore, we present synthetic radio and X-ray maps of the MQR morphology. We aim to identify observational features that can distinguish MQRs from other extended sources, such as supernova remnants or pulsar wind nebulae.

This paper is organized as follows. In Sect.~\ref{sec:model} we describe the physical model for the cocoon. In particular, we analyze the particle injection, transport, and re-acceleration in MQRs. In Sect.~\ref{sec: radiative} we describe all the radiative processes in the system. In Sect.~\ref{sec:results} we present the resulting leptonic distributions, and the radiative signatures for sub-Eddington and super-Eddington systems. In Sect.~\ref{sec:discussion} we discuss the detectability of MQRs. Finally, in Sect.~\ref{sec:conclusions} we summarize the main results and their relevance for the identification of Galactic PeVatrons.

\section{General properties and particle transport}\label{sec:model}

We follow \cite{Abaroa_etal_2026MQR} for the general picture of the MQR. We assume that the MQ remains active for $t_{0} = 10^4\,{\rm yr}$, injecting power through twin jets into the ISM and carving out a dilute bubble that is subsequently filled with relativistic particles accelerated in the shocks formed at the jet termination regions (see e.g., \citealt{Heinz&Sunyaev_2002,2005A&A...432..609B,2009A&A...497..325B, 2010Natur.466..209P,Marti_etal_2015,Marti_etal_2017NatCo, Abaroa_etal2024(S26), Abaroa_etal_2026MQR}). Once the injection ceases, the system evolves into an MQR. 

We adopt a black hole mass of $10\,M_{\odot}$ and study two accretion regimes (for $t < t_0$): a sub-Eddington MQ, with a jet power of $L_{\rm j}=10^{37}\,{\rm erg\, s^{-1}}$, and a super-Eddington MQ with $L_{\rm j}=10^{40}\,{\rm erg\, s^{-1}}$. In both cases we assume a jet Lorentz factor $\gamma_{\rm j}=3$ (e.g., \citealt{Heinz&Sunyaev_2002}), although the model is only weakly sensitive to this parameter.

The semi-major axis of the cocoon equals the jet length during the active MQ phase, $l_{\rm c}(t<t_0)=l_{\rm j}(t<t_0)=(L_{\rm j}/\rho_{\rm ISM})^{1/5}\,t^{3/5}$ \citep{1997MNRAS.286..215K}, while the semi-minor axis at any $t$ is approximated here as $w_{\rm c}\sim l_{\rm c}/3$ \citep{Begelman&Cioffi_1989,2009A&A...497..325B}. Here we assume that the number density of the ISM is roughly the average Galactic disk value, $n_{\rm ISM}=\rho_{\rm ISM}/m_{\rm p}=0.1\, {\rm cm^{-3}}$ (with $m_{\rm p}$ the proton mass). In the MQR phase $(t>t_0)$, the cocoon continues to expand as an adiabatic bubble powered by the previous active phase, treated as an impulsive energy injection. Its 
evolution is approximated as $l_{\rm c}(t>t_0)=(L_{\rm j}\,t_0/\rho_{\rm ISM})^{1/5}\,t^{2/5}$, with a corresponding expansion velocity $v_{\rm c}={\rm d}l_{\rm c}/{\rm d}t$. This adiabatic phase lasts until the onset of the radiative phase (see \citealt{Abaroa_etal_2026MQR}). We evolve the system until the onset of the radiative phase or until the shell expansion becomes dynamically weak and the remnant is expected to merge with the ambient ISM, namely $t_{\rm end}=\min(t_{\rm rad},t_{\rm mix})$ (see \hyperref[app:end_times]{Appendix~\ref{app:end_times}}).

We notice that, for the values of the lifetime of the MQR, the ISM density, and the jet power explored in this work, the expansion velocity of the cocoon is high enough to neglect the proper motion of the system (e.g., $10^7{\rm cm\,s^{-1}}\,10^4\,{\rm yr}\sim 1\,{\rm pc}$). In the following, we describe the properties of the cocoon and study the internal particle transport. Table \ref{tab: parametros generales} lists the main parameters of our fiducial model. In \hyperref[sec: scan]{Sect.~\ref{sec: scan}} we explore additional scenarios varying the jet power, ambient density, and diffusion regime, while \hyperref[appendix]{Appendix~\ref{appendix}} illustrates the effect of diffusion and stochastic re-acceleration on the retained proton population.

\begin{table} 
\begin{center}
\caption{Parameters of the fiducial super- and sub-Eddington scenarios.}
\label{tab: parametros generales}
\begin{adjustbox}{max width=\columnwidth}
\begin{tabular}{l c c c}
\hline
\hline
\rule{0pt}{2.5ex}Parameter & Symbol [units] & Super-Eddington & Sub-Eddington  \\
\hline
%\rule{0pt}{2.5ex}Inclination$^{(1)}$ & $i$ & XX & $^{\circ}$ \\
%\rule{0pt}{2.5ex}Orbital semi-axis$^{(2)}$ & $a$ & $2-13$ & $R_{\odot}$ \\
\rule{0pt}{2.5ex}Age of the MQ$^{(1)}$ & $t_0\,[{\rm yr}]$ & $10^4$ & $10^4$  \\
Lifetime of the MQR$^{(2)}$ & $t_{\rm end}\,[{\rm yr}]$ & $7.5\times10^{4}$ & $3\times 10^{5}$  \\
Jet mechanical power$^{(1)}$ & $L_{\rm j}\,[{\rm erg\,{s}^{-1}}]$  & $10^{40}$ & $10^{37}$ \\
Jet power to relativistic particles$^{(1)}$ & $q_{\rm rel}$  & $0.1$ & $0.1$ \\
Hadron-to-lepton ratio$^{(1)}$ & $K_{\rm ep}$  & $100$ & $100$ \\
Number density of ISM$^{(1)}$ & $n_{\rm ISM}\,[\rm cm^{-3}]$   & $0.1$ & $0.1$   \\
Magnetic field of ISM$^{(1)}$ & $B_{\rm ISM}\,[\mu \rm G]$   & $3$ & $3$   \\
Cocoon magnetic field at $t_0$$^{(2)}$ & $B_{\rm c,0}\,[\mu{\rm G}]$   & $88$ & $22$   \\
Cocoon semi-major axis at $t_0$$^{(2)}$ & $l_{\rm c,0}\,[\rm cm]$   & $7.2\times 10^{19}$ & $1.8\times 10^{19}$   \\
Cocoon expansion speed at $t_0$$^{(2)}$ & $v_{\rm c,0}\,[\rm cm\, s^{-1}]$   & $1.4\times 10^{8}$ & $3.4\times 10^{7}$   \\
Cocoon number density at $t_0$$^{(2)}$ & $n_{\rm c,0}\,[\rm cm^{-3}]$   & $9.6\times10^{-5}$ & $9\times10^{-5}$   \\
\hline
\end{tabular}
\end{adjustbox}
\end{center}
\footnotesize{\textbf{Notes.} We indicate the parameters we have assumed  with superscript ${(1)}$ and those we have derived  with ${(2)}$.
}
\end{table}

\subsection{Evolution of the magnetic field and density in the cocoon}

To estimate the magnetic field in the cocoon, we assume that the magnetic pressure is a fixed fraction of the cocoon pressure,
\begin{equation}
p_B(t)=\eta_B\,p_{\rm c}(t),
\end{equation}
where $\eta_B\sim 0.1$ is a dimensionless parameter. Since $p_B(t)=B^2(t)/8\pi$, the magnetic field is
\begin{equation}
B(t)=\left[8\pi \eta_B p_{\rm c}(t)\right]^{1/2}.
\label{eq:Bofpc}
\end{equation}

For a fluid with effective adiabatic index $\Gamma_{\rm c}$, the cocoon pressure is related to its internal energy density by
\begin{equation}
p_{\rm c}(t)=\left(\Gamma_{\rm c}-1\right)
\frac{E_{\rm c}(t)}{V_{\rm c}(t)},
\label{eq:pc_general_compact}
\end{equation}
where $E_{\rm c}$ and $V_{\rm c}$ are the cocoon internal energy and volume, respectively. The cocoon volume is approximated as $V_{\rm c}(t)\approx 4\pi l_{\rm c}^3(t)/27$, and we adopt $\Gamma_{\rm c}=4/3$, appropriate for a relativistic plasma.

During the active MQ phase, the cocoon is continuously energized by the jets and drives a shock into the surrounding ISM. We normalize the cocoon pressure at the end of the active phase by assuming approximate pressure balance between the cocoon interior and the ram pressure of the swept-up ambient medium,
\begin{equation}
p_{{\rm c},0}\approx \rho_{\rm ISM}v_{{\rm c},0}^2,
\label{eq:pc_t0}
\end{equation}
where $v_{{\rm c},0}$ is the cocoon expansion velocity at $t=t_0$. 

The corresponding internal energy is
\begin{equation}
E_{{\rm c},0}=
\frac{p_{{\rm c},0}V_{{\rm c},0}}{\Gamma_{\rm c}-1}.
\label{eq:Ec_t0}
\end{equation}
For the adopted active-phase scaling, $l_{\rm c}\propto (L_{\rm j}/\rho_{\rm ISM})^{1/5}t^{3/5}$, this estimate gives
\begin{equation}
E_{{\rm c},0}\sim 0.5\,L_{\rm j}t_0,
\end{equation}
consistent with the expectation that a substantial fraction of the injected jet energy remains in the shocked cocoon, while the rest is transferred to the swept-up shell and to mechanical work. The corresponding magnetic field at $t=t_0$ is
\begin{equation}
B_{{\rm c},0}=
\left(8\pi\eta_B p_{{\rm c},0}\right)^{1/2}.
\label{eq:Bc_t0}
\end{equation}
The lower limit for the field is assumed to be that of the ISM,
$B_{\rm c}^{\rm min}=B_{\rm ISM}$, which is an ansatz.

We also estimate the mean density of the cocoon from the mass contained in it at the end of the active phase. As we describe in the next subsection, the cocoon density is a relevant parameter for the re-acceleration process within the cocoon. We write $\rho_{{\rm c},0}=M_{{\rm c},0}/V_{{\rm c},0}$, where $M_{{\rm c},0}$ includes both the mass injected by the jet and a small fraction of ambient material mixed into the cocoon. The jet contribution is estimated from its kinetic power, assuming a cold jet, as
\begin{equation}
\dot{M}_{\rm j}=
\frac{L_{\rm j}}{(\gamma_{\rm j}-1)c^2},
\label{eq:mdot_jet}
\end{equation}
so that $M_{{\rm j},0}\sim\dot{M}_{\rm j}t_0$. In addition, hydrodynamical simulations of jet--ISM interactions show that the cocoon can contain mixed material entrained from the shocked ambient medium \citep{Bosch-Ramon_etal2011A&A}. We therefore include an external mixing term,
\begin{equation}
M_{\rm mix}=f_{\rm mix}M_{\rm sw},
\label{eq:Mmix}
\end{equation}
where $f_{\rm mix}\ll 1$ is the fraction of swept-up material that becomes mixed into the cocoon, and
\begin{equation}
M_{\rm sw}\sim \rho_{\rm ISM}\frac{4\pi}{3}l_{{\rm c},0}^3
\label{eq:Mswept}
\end{equation}
is an approximate estimate of the swept-up mass. The total cocoon mass at jet shutdown is then $M_{{\rm c},0}=M_{{\rm j},0}+M_{\rm mix}$. This prescription treats the density obtained from the jet mass flux as a lower limit, while allowing for a small degree of turbulent mixing with the shocked ISM. Here we assume $f_{\rm mix}\sim10^{-4}$.

After jet shutdown $(t>t_0)$, the cocoon is no longer replenished by the jet, but it continues to expand as an energy-driven remnant. In our one-zone description, we keep the cocoon pressure tied to the ram pressure of the swept-up ambient medium,
\begin{equation}
p_{\rm c}(t)=p_{{\rm c},0}
\left[\frac{v_{\rm c}(t)}{v_{{\rm c},0}}\right]^2.
\label{eq:pc_evolution_dynamic}
\end{equation}
For the adopted remnant-phase expansion, $(l_{\rm c}\propto t^{2/5})$, the expansion velocity scales as $v_{\rm c}\propto t^{-3/5}$. Therefore,
\begin{equation}
p_{\rm c}(t)\propto t^{-6/5}.
\end{equation}
Since the magnetic pressure is assumed to remain a fixed fraction of the cocoon pressure, the magnetic field evolves as
\begin{equation}
B_{\rm c}(t)=
\max\left[
B_{\rm ISM},
B_{{\rm c},0}
\frac{v_{\rm c}(t)}{v_{{\rm c},0}}
\right],
\qquad (t>t_0).
\label{eq:B_piecewise_dynamic}
\end{equation}
Equivalently, for $l_{\rm c}\propto t^{2/5}$,
\begin{equation}
B_{\rm c}(t)=
\max\left[
B_{\rm ISM},
B_{{\rm c},0}
\left(\frac{t}{t_0}\right)^{-3/5}
\right],
\qquad (t>t_0).
\label{eq:B_piecewise_compact}
\end{equation}

For the density, we assume that no additional mass is loaded into the cocoon after the jet turns off, so that $M_{\rm c}(t>t_0)\simeq M_{{\rm c},0}$. Therefore, $\rho_{\rm c}(t)\propto V_{\rm c}(t)^{-1}$. Since $V_{\rm c}\propto l_{\rm c}^3\propto t^{6/5}$ in the remnant phase, the mean cocoon density evolves as
\begin{equation}
\rho_{\rm c}(t)=
\rho_{{\rm c},0}\left(\frac{t}{t_0}\right)^{-6/5},
\qquad (t>t_0).
\label{eq:rho_piecewise_compact}
\end{equation}

These prescriptions provide a phenomenological description of the cocoon evolution in which the internal pressure remains linked to the ram pressure of the swept-up ISM, while the magnetic pressure is kept as a fixed fraction of the cocoon pressure. The density evolution follows from the expansion of a fixed cocoon mass after jet shutdown. This approach is intended to maintain consistency between the adopted expansion law and the pressure that drives the external shock.

\subsection{Particle transport and distribution}\label{sect: cocoon}

Particles accelerated at the jet termination region of the MQ during the active phase $(t<t_0)$ are advected by the bowshock backflows and subsequently injected into the cocoon (e.g., \citealt{Heinz&Sunyaev_2002,Abaroa_etal2024(S26)}).  
Escape is assumed to be diffusive in the cocoon:
\begin{equation}
    t_{\rm esc,c}(E,t)=\frac{\Delta x_{\rm c}(t)^2}{6D(E,t)},
    \label{eq:tesc_cocoon}
\end{equation}
where $\Delta x_{\rm c}(t) \sim l_{\rm c}(t)$ is the characteristic escape length and $D(E,t)$ is the spatial diffusion coefficient. We parameterize the latter as \citep{Ptuskin2012,Peretti2025}
\begin{equation}
    D(E,t) \approx \frac{1}{3}\,c\,L_{\mathrm{c}}(t)
    \left( \frac{r_{\mathrm{L}}(E,t)}{L_{\mathrm{c}}(t)} \right)^{\delta}
    \left( \frac{B(t)}{\Delta B(t)} \right)^2 .
    \label{eq:spatial_diffusion}
\end{equation}
Here, $r_{\mathrm{L}}(E,t)=E/eB(t)$ is the Larmor radius of a relativistic particle with energy $E$, $B$ is the mean magnetic field, $\Delta B$ its turbulent component, and $L_{\mathrm{c}}(t)$ is the coherence or injection scale of the turbulence, where most of the magnetic energy is contained. Here we assume in all cases that $\Delta B / B= 0.9$. This choice corresponds to magnetic
fluctuations comparable to the mean field and therefore to efficient particle
scattering.

In contrast to our previous work \citep{Abaroa_etal_2026MQR}, where particle
transport was modeled close to the Bohm limit ($\delta=1$) to maximize
confinement, here we adopt a Kolmogorov diffusion regime ($\delta=1/3$) as our
fiducial case. This choice is motivated by the expectation that, after jet
shutdown, the cocoon contains a developed turbulent cascade generated by jet
backflows, internal shocks, and jet--medium mixing. A similar assumption has
been adopted for particle transport in wind-blown bubbles around massive stellar
clusters, where Kolmogorov turbulence leads to a diffusion coefficient scaling
as $D(E)\propto E^{1/3}$ \citep{Morlino_etal_2021}. Hydrodynamical simulations of
wind-driven bubbles also suggest that stellar feedback can generate
Kolmogorov-like turbulence in hot cavities \citep{Gallegos_Garcia_2020ApJ}. We
therefore regard Kolmogorov diffusion as a representative case for transport in
the turbulent MQR cocoon. We set $L_{\mathrm{c}}(t)=\chi_{\rm turb}l_{\rm c}(t)$, with $\chi_{\rm turb}=10^{-2}$, corresponding to a small fraction of the cocoon size and to the expected correlation length of turbulence within the cocoon (see, e.g., \citealt{Zhang_etal_2025}).

After jet shutdown, the turbulent motions generated during the active phase are no longer continuously driven. We therefore assume that the turbulent power available for stochastic re-acceleration decreases during the remnant phase. Since the detailed decay of the turbulent cascade is not resolved in our one-zone model, we introduce a phenomenological attenuation factor,
\begin{equation}
    f_{\rm turb}(t)=
    \begin{cases}
    1, & t\leq t_0,\\[0.2cm]
    \exp\left[-(t-t_0)/\tau_{\rm turb}\right], & t>t_0,
    \end{cases}
    \label{eq:fturb}
\end{equation}
where $\tau_{\rm turb}$ is an effective turbulence-decay timescale. Here we assume $\tau_{\rm turb}=t_0/2$. 

We incorporate stochastic re-acceleration in the cocoon, driven by the remaining internal turbulence and modeled as a second-order Fermi process. The acceleration rate before turbulent attenuation is given by (e.g., \citealt{Romero&Muller&Roth2018Starburst})
\begin{equation}
    t^{-1}_{\rm acc,0}(E,t)=\frac{v_{\rm A}^2(t)}{D(E,t)}
    \approx 3\frac{c}{L_{\rm c}(t)}
    \left(\frac{v_{\rm A}(t)}{c}\right)^2
    \left(\frac{\Delta B(t)}{B(t)}\right)^2
    \left(\frac{L_{\rm c}(t)}{r_{\rm L}(E,t)}\right)^\delta ,
    \label{eq:tacc_fermiII}
\end{equation}
where $v_{\rm A}(t)=B(t)/\sqrt{4\pi \rho_{\rm c}(t)}$ is the Alfvén velocity.

We solve the time-dependent transport equation to determine the evolution of the relativistic particles injected in the cocoon from the reverse shock of the jet. For each particle species, the transport equation can be written as (see \citealt{Romero&Muller&Roth2018Starburst,Abaroa_etal_2026MQR}, and references therein)  
\begin{dmath}\label{ec:transportDSA}
   \frac{\partial n(E,t)}{\partial t}
   =
   Q(E,t)
   -\frac{\partial}{\partial E}
   \left[\dot{E}(E,t)n(E,t)\right]
   +\frac{\partial}{\partial E}
   \left[
   K(E,t)\frac{\partial n(E,t)}{\partial E}
   -
   A(E,t)n(E,t)
   \right]
   -\frac{n(E,t)}{t_{\rm esc}(E,t)} .
\end{dmath}
Here, $n(E,t)$ is the differential particle density and $\dot{E}(E,t)$ accounts for deterministic energy losses. The injection term, \(Q(E,t)\), accounts for particles accelerated at the jet termination region and subsequently transported into the larger, more dilute cocoon, including the rapid adiabatic cooling and spatial dilution associated to the transport \citep{Abaroa_etal_2026MQR}.

The third term on the right-hand side of Eq.~(\ref{ec:transportDSA}) corresponds to stochastic re-acceleration. The energy diffusion coefficient is
\begin{equation}
    K(E,t)=\frac{v_{\rm A}^2(t)}{D(E,t)}E^2 f_{\rm turb}(t),
    \label{eq:K_energy_diffusion}
\end{equation}
which characterizes the diffusion of particles in energy space. The associated systematic drift coefficient is
\begin{equation}
    A(E,t)=\frac{2K(E,t)}{E}.
    \label{eq:A_drift}
\end{equation}

To maintain energetic consistency, we monitor the instantaneous energy gain produced by the stochastic term and require the corresponding power to remain below a prescribed fraction of the available non-thermal power. For electrons and protons, respectively, the available powers are
\begin{equation}
    L_{\rm e}=\frac{q_{\rm rel}}{K_{\rm ep}+1}L_{\rm j},
    \qquad
    L_{\rm p}=\frac{K_{\rm ep}q_{\rm rel}}{K_{\rm ep}+1}L_{\rm j},
    \label{eq:Le_Lp_available}
\end{equation}
where $K_{\rm ep}=L_{\rm p}/L_{\rm e}$ is the hadron-to-lepton energy ratio. We impose that the non-thermal power that can be supplied through stochastic re-acceleration is $\leq \xi_{\rm FII,i} L_i\ \ (i={\rm e,p})$, where $\xi_{\rm FII,i}\lesssim 1$. This condition prevents Fermi-II acceleration from acting as an unbounded energy source in the remnant phase.

As mentioned above, in the remnant phase the cocoon expands as an adiabatic bubble powered by an impulsive energy input, with $v_{\rm c}\propto t^{-3/5}$. The deceleration of the shell is therefore gradual. A simple comparison between the sound-crossing time of the shocked shell and the deceleration time shows that the shell can adjust to the evolving flow on a timescale shorter than the dynamical time (see Appendix \ref{appendix_shock}).

\section{Radiative processes and synthetic maps}\label{sec: radiative}

We describe here the different radiative components of the MQR, i.e., the thermal and non-thermal processes that lead to the emission of the cocoon and its shell. We also describe the synthetic radio and X-ray maps we construct to analyze the response of the instruments to MQRs.

\subsection{Non-thermal emission of the cocoon}

Particle cooling in the presence of ambient fields causes the MQR to emit non-thermal radiation. Due to the very low density of the cocoon (significantly lower than the typical ISM density), relativistic protons are expected to cool inefficiently inside the MQR, even over long timescales \citep{Abaroa_etal_2026MQR}. Electrons, on the contrary, efficiently interact with the intrinsic magnetic field of the cocoon through synchrotron radiation, while inverse Compton scattering occurs on the cosmic microwave background ($T_{\rm CMB}=2.725\,$K, $u_{{\rm CMB}}=4.2\times 10^{-13}\,{\rm erg\,cm^{-3}}$) and  the Galactic infrared radiation field ($T_{\rm IR}\approx 30\,$K, $u_{{\rm IR}}\approx 10^{-12}\,{\rm erg\,cm^{-3}}$).  
Synchrotron self-Compton (SSC) may also be relevant in some scenarios. Both particle populations ---hadrons and leptons--- also experience adiabatic losses due to the work performed during the cocoon expansion. Synchrotron self-absorption (SSA) is included, although it only affects the spectra at very low radio frequencies.

We therefore calculate the synchrotron, inverse Compton, and SSC emission produced by relativistic leptons confined in the cocoon in order to characterize its radiative output. Details of these calculations can be found in, e.g., \cite{Bosch-Ramon&Khangulyan_2009_MQreview}, \cite{2011hea..book.....L}, and \cite{Romero-Vila2014}.

\subsection{Non-thermal emission of the cocoon's shell}

Relativistic protons originally injected by the jet and subsequently escaping from the cocoon interact with the swept-up shell material. We model the shell as a prolate layer surrounding the cocoon and assume that it has the same aspect ratio as the cocoon. The semi-major axis of the inner boundary is equal to the semi-major axis of the cocoon, $l_{\rm in}(t)=l_{\rm c}(t)$, whereas the outer boundary is $l_{\rm out}(t)=l_{\rm c}(t)+\Delta l_{\rm c}(t)$. The shell volume is computed as the difference between the outer and inner ellipsoids,
\begin{equation}
    V_{\rm sh}(t)=
    \frac{4\pi}{3\mathcal{A}^2}
    \left[
    l_{\rm out}^3(t)-l_{\rm c}^3(t)
    \right],
    \label{eq:V_shell}
\end{equation}
where $\mathcal{A}\equiv l_{\rm c}/w_{\rm c}=3$ is the adopted aspect ratio.

The shell density and thickness are linked by mass conservation. We write the mean shell density as $\rho_{\rm sh}=\kappa_{\rm sh}\rho_{\rm ISM}$, where \(\kappa_{\rm sh}\) is the mean compression factor. Requiring the swept-up ISM mass to be contained in the shell gives
\begin{equation}
    \rho_{\rm ISM}V_{\rm out}
    =
    \rho_{\rm sh}V_{\rm sh},
\end{equation}
or equivalently
\begin{equation}
    V_{\rm out}
    =
    \kappa_{\rm sh}
    \left(
    V_{\rm out}-V_{\rm in}
    \right).
\end{equation}
Since both the inner and outer boundaries are assumed to have the same aspect ratio, \(V\propto l^3\), and therefore
\begin{equation}
    \frac{l_{\rm out}}{l_{\rm c}}
    =
    \left(
    \frac{\kappa_{\rm sh}}{\kappa_{\rm sh}-1}
    \right)^{1/3}.
    \label{eq:lout_lc_shell}
\end{equation}
For a strong adiabatic shock, \(\kappa_{\rm sh}=4\), which gives
\begin{equation}
    \frac{l_{\rm out}}{l_{\rm c}}
    =
    \left(\frac{4}{3}\right)^{1/3}
    \simeq 1.10,
\end{equation}
and hence
\begin{equation}
    \Delta l_{\rm c}(t)
    =
    l_{\rm out}(t)-l_{\rm c}(t)
    \simeq 0.1\,l_{\rm c}(t).
\end{equation}
Thus, the adopted shell thickness and density are not independent, but are linked through mass conservation. For simplicity, we take \(B_{\rm sh}=4B_{\rm ISM}\), noting that the exact magnetic-field amplification depends on the upstream field orientation and turbulence.

Particles escaping from the cocoon are not assumed to interact instantaneously with the shell. Instead, we treat their propagation across the swept-up material as diffusive, using the same diffusion prescription adopted for the cocoon but evaluated with the physical parameters of the shell. The characteristic residence time in the shell is therefore estimated as $t_{\rm esc,sh}(E,t) = \Delta l_{\rm c}^2(t)/6D_{\rm sh}(E,t)$, where $\Delta l_{\rm c}(t)$ is the shell thickness and $D_{\rm sh}(E,t)$ is calculated using the compressed magnetic field.  During this residence time, relativistic protons interact with the swept-up gas through $pp$ collisions, while primary and secondary electrons radiate through synchrotron, relativistic Bremsstrahlung, and inverse Compton processes. In the $pp$ process, along with $\pi^0$ decay, electron-positron pairs are created and interact with ambient fields contributing to the non-thermal emission of the shell. In addition, we estimate absorption at very low energies through SSA.

\subsection{Thermal emission of the cocoon's shell}

In addition to non-thermal emission from relativistic particles, the expansion of the cocoon produces a shocked shell that can radiate thermally. Therefore, we complement our calculations for the non-thermal emission of the MQR with an estimate of the thermal radiation produced by the shocked ISM as the cocoon expands. Since the swept-up shell is expected to be optically thin, we model the emission as thermal Bremsstrahlung radiation from a collisionally ionized plasma.

The emissivity of a thermal, optically-thin plasma at a temperature $T$ for a frequency $\nu$, is given by \citep{2011hea..book.....L}:
\begin{dmath}\label{eq: emisssivity}
    j_{\nu}[T(t)]= \mathbb{C}\,Z^2\, T(t)^{-1/2}\,n_{\rm i}\, n_{\rm e}\, 
    g(\nu,T) \times \, \exp{\left(\frac{-h\nu}{k_{\rm B}T(t)}\right)}\ {\rm erg\,s^{-1}\,cm^{-3}\,Hz^{-1}},
\end{dmath}
where $\mathbb{C}=6.8\times10^{-38}$, $Z$ is the atomic number, $n_{\rm i}$ and $n_{\rm e}$ are the number density of ions and electrons, $h$ is the Planck constant, and $k_{\rm B}$ the Boltzmann constant. For a fully ionized hydrogen plasma, $n_{\rm e}\simeq n_{\rm i}\simeq n_{\rm sh}$, with $n_{\rm sh}$ the post-shock density. The Gaunt factor is approximated here as $g(\nu,T)\sim 1$. The temperature in Eq. \eqref{eq: emisssivity} is in our case the post-shock temperature of the swept-up shell, which is set by the cocoon expansion speed:
\begin{equation}
T_{\rm sh}(t)=\frac{3\mu m_{\rm p}}{16k_{\rm B}}\,v_{\rm c}^2(t),
\label{eq:Tps}
\end{equation}
where $\mu=0.5$ is the mean molecular weight.

\subsection{Surface brightness of the microquasar remnant}

Unlike compact sources, MQRs are expected to be large, diffuse, and approximately homogeneous emitters. As a consequence, their detectability is not primarily determined by the total luminosity or the spectral energy distribution (SED), but rather by the surface brightness of the emission.

Extended sources with low surface brightness may remain undetected even if their integrated luminosity is significant, since most high-energy instruments are limited by background and angular resolution \citep{Combi1998A&A,Abaroa_etal_2026MQR}. For this reason, we focus on the surface brightness as the relevant observable quantity to assess the detectability of MQRs at different wavelengths.

We first calculate the luminosity for each radiative process and then assume a
fiducial Galactic distance $d=2\,{\rm kpc}$ to estimate the corresponding flux.
The surface brightness is estimated as
\begin{equation}
    \Sigma(t) =
    \frac{L(t)}{4\pi d^2\,\Omega(t)} .
\end{equation}
For each emitting component, the cocoon and the shell, we estimate the
corresponding projected solid angle, $\Omega_{\rm c,sh}(t)$.

Since the cocoon is modeled as a prolate structure, with semi-major axis
$l_{\rm c}(t)$ and semi-minor axis $w_{\rm c}(t)=l_{\rm c}(t)/3$, we compute
the solid angle from its projected area.
For an inclination angle $i$ between the cocoon axis and the line of sight, the
projected semi-major axis is approximated as
\begin{equation}
    a_{\rm proj}(t) =
    \left[
    l_{\rm c}^2(t)\sin^2 i +
    w_{\rm c}^2(t)\cos^2 i
    \right]^{1/2},
\end{equation}
whereas the projected semi-minor axis is $b_{\rm proj}(t)=w_{\rm c}(t)$.
The cocoon solid angle is therefore
\begin{equation}
    \Omega_{\rm c}(t) \simeq
    \frac{\pi a_{\rm proj}(t)b_{\rm proj}(t)}{d^2}.
\end{equation}
For an edge-on configuration, $i=90^\circ$, this
reduces to
\begin{equation}
    \Omega_{\rm c}(t) \simeq
    \frac{\pi l_{\rm c}(t)w_{\rm c}(t)}{d^2}.
\end{equation}

In the case of the shell, we estimate the solid angle as the difference between the
projected areas of the outer and inner ellipses. This choice approximates the solid angle of the limb-brightened part of the shell. Although an optically thin shell also contributes emission inside the inner projected ellipse, the largest line-of-sight path lengths, and hence the highest surface brightness, occur near the projected rim. We take the inner semi-major
axis to be $l_{\rm in}(t)=l_{\rm c}(t)$ and the outer one to be
$l_{\rm out}(t)=l_{\rm c}(t)+\Delta l_{\rm c}(t)$, with the same aspect ratio
as the cocoon. Thus,
\begin{equation}
    \Omega_{\rm sh}(t) \simeq
    \frac{\pi}{d^2}
    \left[
    a_{\rm out,proj}(t)b_{\rm out,proj}(t)
    -
    a_{\rm in,proj}(t)b_{\rm in,proj}(t)
    \right],
\end{equation}
where the projected semi-axes are computed as above. For $i=90^\circ$, this
becomes
\begin{equation}
    \Omega_{\rm sh}(t) \simeq
    \frac{\pi}{d^2}
    \left[
    l_{\rm out}(t)w_{\rm out}(t)
    -
    l_{\rm in}(t)w_{\rm in}(t)
    \right].
\end{equation}

\subsection{Radio and X-ray synthetic maps}

To connect the model predictions with realistic observational strategies, we also compute instrument-oriented synthetic maps in radio and soft X-rays  (see \hyperref[app: maps]{Appendix~\ref{app: maps}}, for more details). We construct the maps for an inclination angle of $90^\circ$, and we notice that changes in the inclination lead to changes in the morphology of the MQR. 

In the radio band, we focus on MeerKAT-like observations at $\nu=1.3$ GHz, motivated by the MeerKAT Galactic Plane Survey \citep{Meerkat2024}. Although MeerKAT provides arcsecond-scale angular resolution, the MQR emission predicted by our model is highly extended and of low surface brightness. We therefore smooth the intrinsic radio maps to an effective beam of $5'$ FWHM, representative of an imaging strategy optimized for diffuse emission rather than compact structures. The intrinsic specific-intensity map, $I_\nu(x,y)$, is convolved with a Gaussian beam, and then converted into flux density per beam according to $S_{\nu,\rm beam}(x,y) = I_{\nu,\rm conv}(x,y)\,\Omega_{\rm beam}$, where
\begin{equation}
    \Omega_{\rm beam} =
    \frac{\pi}{4\ln 2}\,\theta_{\rm beam}^2
\end{equation}
for a circular Gaussian beam. The resulting radio maps are shown in units of $\mu{\rm Jy\,beam^{-1}}$ (see Sect. \ref{sec: synthetic}). We stress that these beam-smoothed maps assume full recovery of the large-scale emission and therefore represent optimistic estimates. We refer to \cite{Thompson2017} for details on construction of synthetic radio maps. 

For the soft X-ray band, we generate \textit{eROSITA}-like maps in the $0.5$--$2$ keV range. This band is well suited to the thermal and non-thermal X-ray components expected from the shocked shell and cocoon, and \textit{eROSITA} is particularly relevant because of its wide-field survey capability \citep{eRositaBook2012}. Starting from the intrinsic band-integrated intensity map, $I_X(x,y)$, we compute the expected number of counts per pixel as
\begin{equation}
    N_{\rm ct}(x,y) =
    I_X(x,y)\,\Omega_{\rm pix}\,
    A_{\rm eff}\,t_{\rm exp}\,
    \frac{\mathcal{T}_{X}}{\langle E_\gamma\rangle},
\end{equation}
where $\Omega_{\rm pix}$ is the pixel solid angle, $A_{\rm eff}$ is the effective area, $t_{\rm exp}$ is the exposure time, $\langle E_\gamma\rangle$ is the representative photon energy, and $\mathcal{T}_{X}$ is an effective band-averaged transmission factor accounting for interstellar absorption (here we assumed $\mathcal{T}_{X}=1$). We adopt a Gaussian approximation to the \textit{eROSITA} point-spread function with FWHM $\sim30''$. We notice that these maps provide a first-order estimate of the expected X-ray count distribution; a detailed comparison with a particular observation would require the full instrumental response, spatially varying exposure, realistic background, and line-of-sight absorption.

\section{Results}\label{sec:results}

In this section, we present the results for the electron distribution in the cocoon, the radiative outputs of MQRs in terms of their SEDs and surface brightness, and the synthetic maps. We then present some results obtained by varying the ISM density and diffusion regime of our fiducial models. We focus on the temporal and spectral evolution of the radiative processes described in the previous section, and analyze how the extended and diffuse nature of the source shapes its observational appearance. This approach allows us to directly evaluate the detectability of MQRs at different wavelengths as a function of their age and physical size. We notice that, for the sub-Eddington scenario, the hydrodynamical lifetime can extend up to \(\sim 3\times10^5\,\mathrm{yr}\), but we show the radiative evolution up to \(10^5\,\mathrm{yr}\), by which time the intrinsic emission has already become very faint in the bands considered here.

\subsection{Electron distributions}

\begin{figure}
  \centering
    \centering
    \includegraphics[width=\columnwidth]{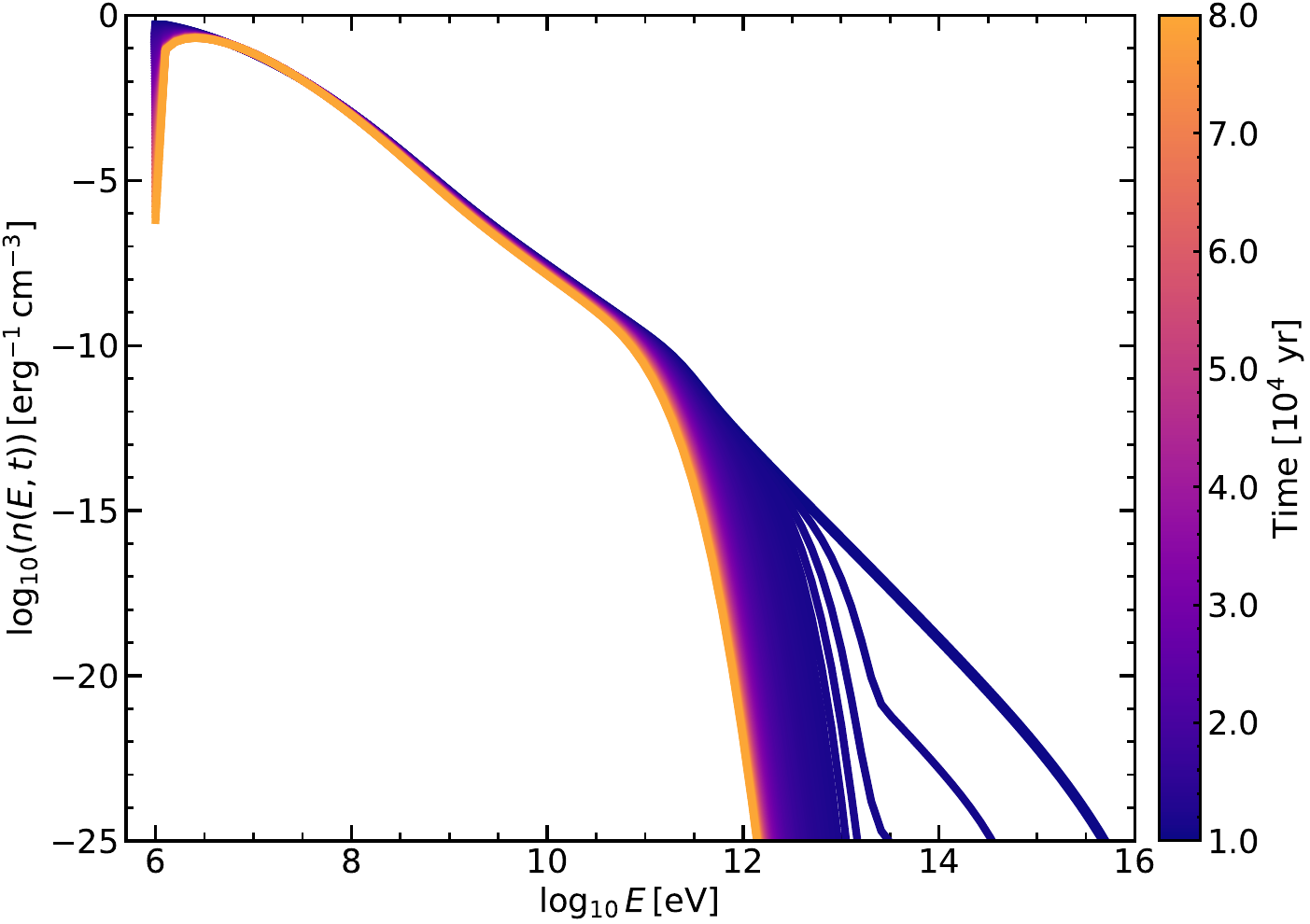}
    \centering
    \includegraphics[width=\columnwidth]{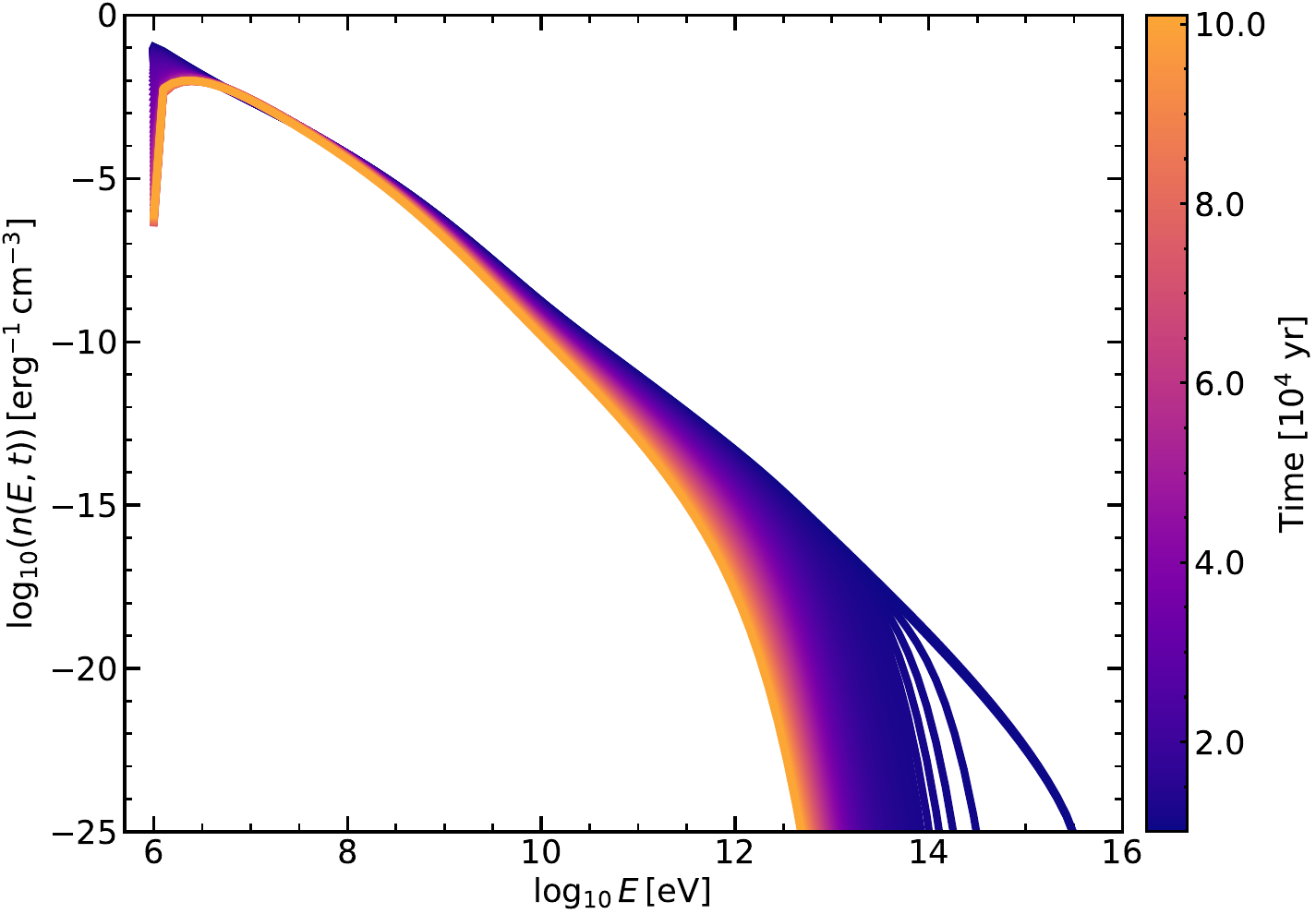}
  \caption{Evolution of the electron distributions in the cocoon over the lifetime of the MQR. Top: super-Eddington system; bottom: sub-Eddington system.}
  \label{fig: distribution}
\end{figure}

We solve the transport equation (Eq. \ref{ec:transportDSA}) to obtain the evolution of the electron distribution in the cocoon over the lifetime of the MQR. \hyperref[fig: distribution]{Figure~\ref{fig: distribution}} shows the results for the super-Eddington (top) and sub-Eddington (bottom) regimes.

The inclusion of stochastic re-acceleration modifies the temporal evolution of the electron distribution by redistributing particles in energy space. Nevertheless, radiative losses, adiabatic expansion, and escape continue to reduce the overall normalization of the distribution as the cocoon evolves.

The ankle-like feature visible in the early super-Eddington curves is associated with the rapid reshaping of the high-energy electron population shortly after jet shutdown. At these times, the distribution still retains a high-energy tail inherited from the active MQ phase, while radiative losses, adiabatic dilution, escape, and the gradual decay of stochastic re-acceleration act to erode this tail. The feature therefore marks the transition between the lower-energy part of the distribution, which evolves more gradually, and the high-energy part, which cools and escapes more efficiently. As the remnant evolves, this transient break is smoothed out and the high-energy cutoff moves to lower energies.

In \hyperref[appendix]{Appendix~\ref{appendix}}, we compare the fiducial models with Kraichnan diffusion and with a Bohm-like limiting case, in order to illustrate the sensitivity of the retained proton population to the degree of confinement. We also compare, for each case, the proton distribution with and without the re-acceleration process.

\subsection{Spectral energy distributions and surface brightness}

\begin{figure}
  \centering
    \centering
    \includegraphics[width=\columnwidth]{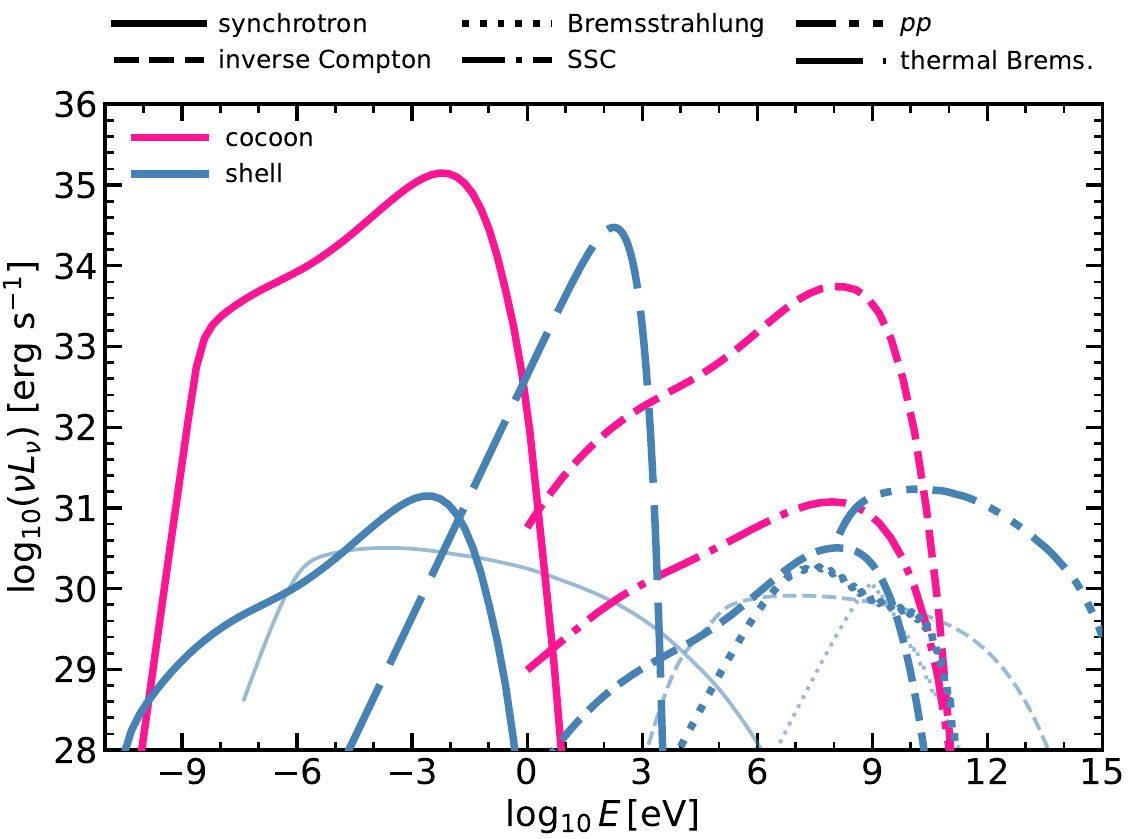}
    \centering
    \includegraphics[width=\columnwidth]{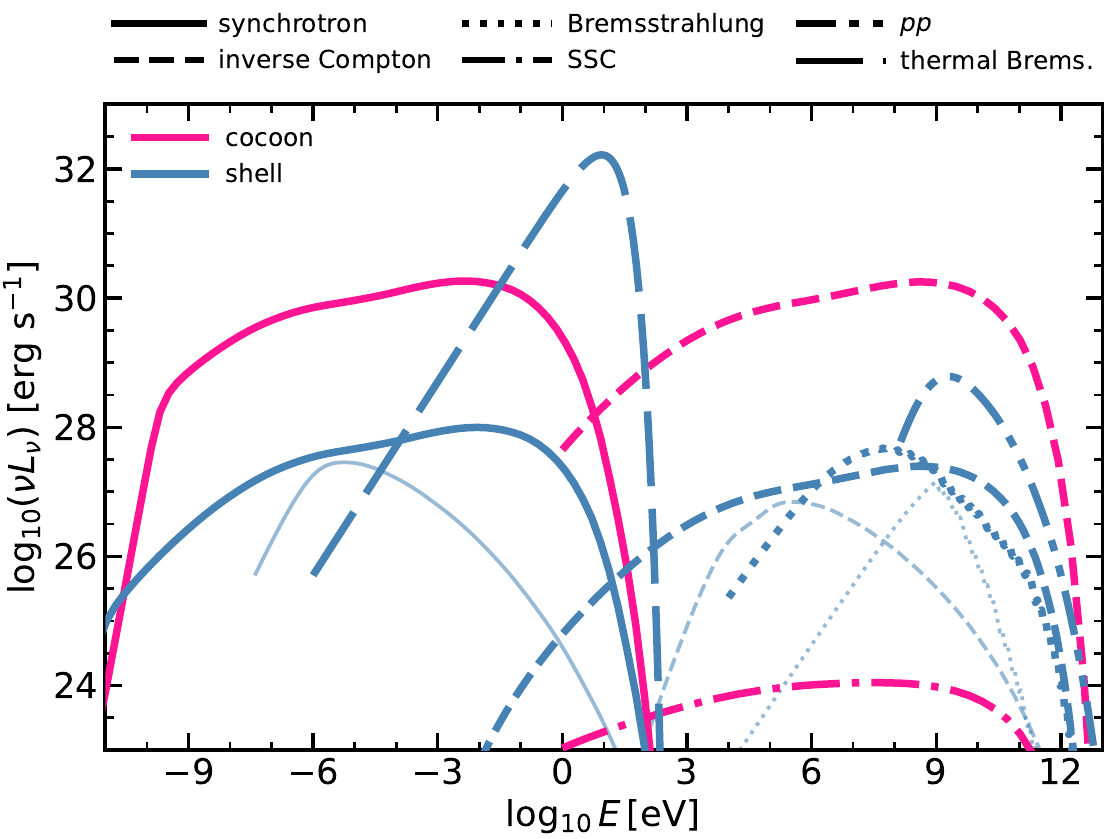}
  \caption{\small  Spectral energy distributions at one time: $t=4\times10^{4}\,$yr for the super-Eddington system (top) and $t=5\times10^{4}\,$yr for the sub-Eddington system (bottom). Colors distinguish the emitting regions: pink for the cocoon and blue for the shell. Line styles distinguish the different radiative processes. Thin lines show the contribution from secondary pairs produced in hadronic interactions in the shell.} 
  \label{fig: SED_onetime}
\end{figure}

\hyperref[fig: SED_onetime]{Figure~\ref{fig: SED_onetime}} shows the SEDs of the different radiative thermal and non-thermal processes considered, again for the super- and sub-Eddington regimes. We show results for an intermediate epoch of the MQR lifetime: $t=4\times10^{4}\,$yr and $t=5\times10^{4}\,$yr, respectively. We note that the SEDs shown here are intrinsic to the source: we include only absorption processes internal to the MQR and do not account for foreground attenuation by the Galactic ISM.

The maximum luminosity in the super-Eddington case is given by the synchrotron radiation from the cocoon, which reaches $\gtrsim 10^{35}\,{\rm erg\,s^{-1}}$. In the sub-Eddington case, the maximum luminosity is given by the thermal emission of the shell, at soft X-ray energies, and reaches $\gtrsim 10^{32}\,{\rm erg\,s^{-1}}$. The non-thermal emission is dominated by the synchrotron and inverse Compton processes in the cocoon in both scenarios, although Bremsstrahlung is also relevant in the sub-Eddington case. Furthermore, in the super-Eddington scenario, the $pp$ process is relevant at VHE energies. The contribution of the secondary pairs (shown as thin lines in \hyperref[fig: SED_onetime]{Fig,~\ref{fig: SED_onetime}}) created in the $pp$ interaction of protons escaping through the shell is not relevant in any case. The SSC process is negligible in the sub-Eddington scenario, while in the super-Eddington case it is comparable to the synchrotron and $pp$ in the shell.

\begin{figure}  
  \centering
    \centering
    \includegraphics[width=\columnwidth]{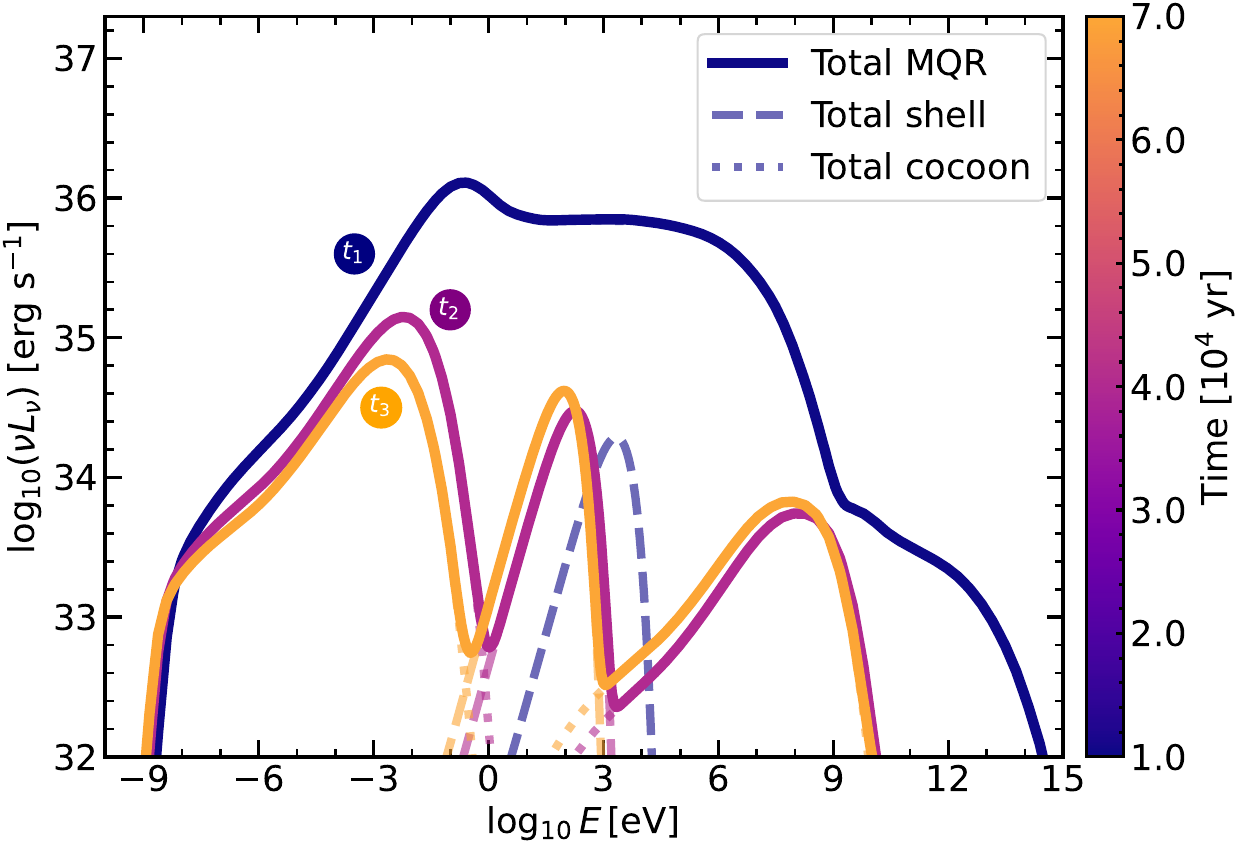}
    \centering
    \includegraphics[width=\columnwidth]{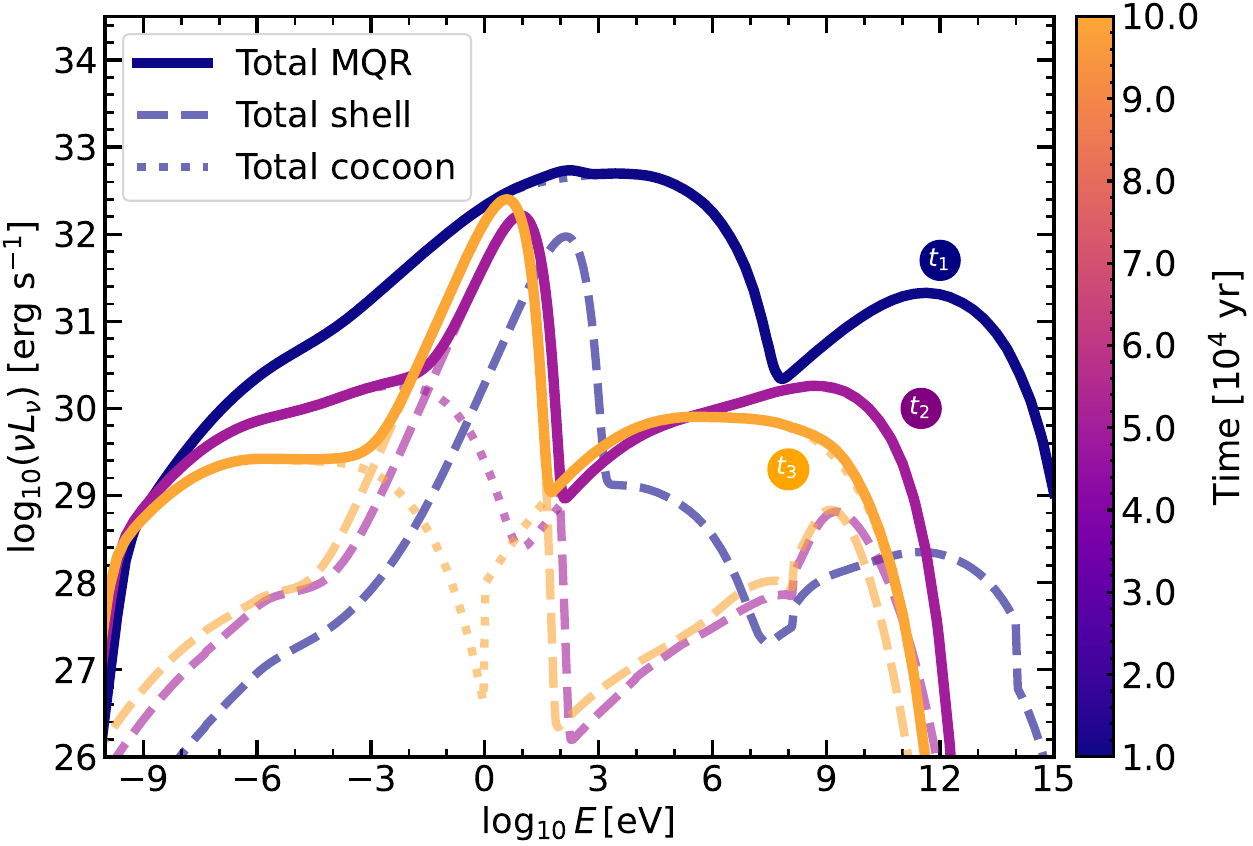}
  \caption{Total SEDs for the super-Eddington (top) and sub-Eddington (bottom) scenarios at three different epochs ($t_{1,2,3}$). Solid lines show the total MQR emission, defined as the sum of the cocoon and shell contributions. Dashed and dotted lines show the shell and cocoon contributions separately.}
  \label{fig: SED_total}
\end{figure}

Once we have calculated the individual components of the SEDs, we sum all contributions for each of the three selected times $(t_{1,2,3})$, corresponding to initial (blue), intermediate (purple), and final (yellow) epochs of the MQR during its lifetime. In \hyperref[fig: SED_total]{Fig.~\ref{fig: SED_total}} we show the total SEDs for the super-Eddington (top) and sub-Eddington (bottom) scenarios. 

It can be seen that, in all non-thermal processes, the emission decreases as the system evolves. On the contrary, the thermal emission shifts to lower energies but slightly increases its associated luminosity. The luminosity of the system evolves rapidly at the initial epoch of the MQR, decreasing by orders of magnitude across the spectrum (with the exception of the soft X-rays), in agreement with the results on particle distribution (\hyperref[fig: distribution]{Fig.~\ref{fig: distribution}}). We notice that the difference between the luminosities obtained in each scenario  roughly follows the contrast in injected jet power ($L_{\rm jet,super}/L_{\rm jet,sub}=10^{40}/10^{37}=10^3$).

\begin{figure}
  \centering
    \centering
    \includegraphics[width=\columnwidth]{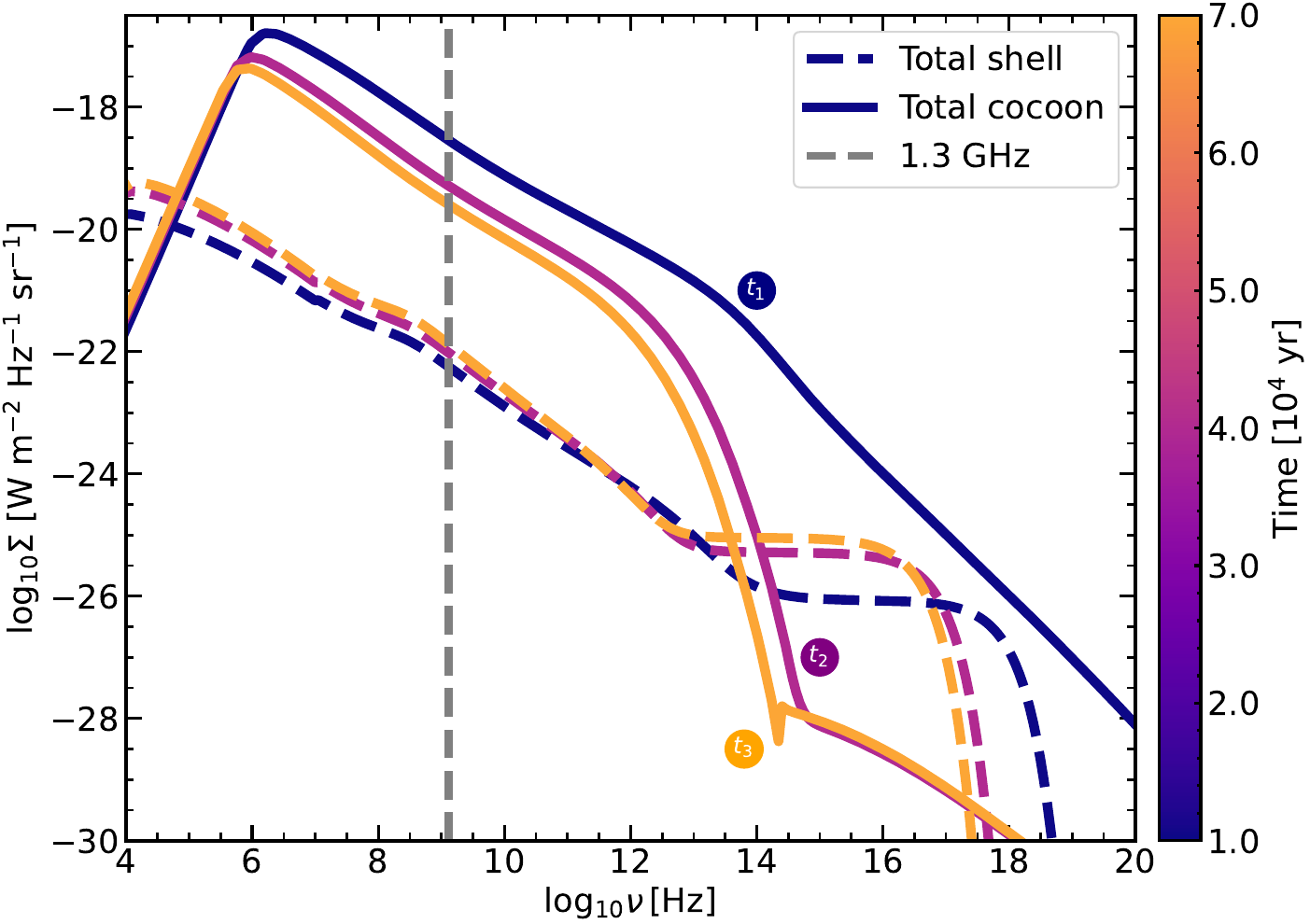}
    \centering
    \includegraphics[width=\columnwidth]{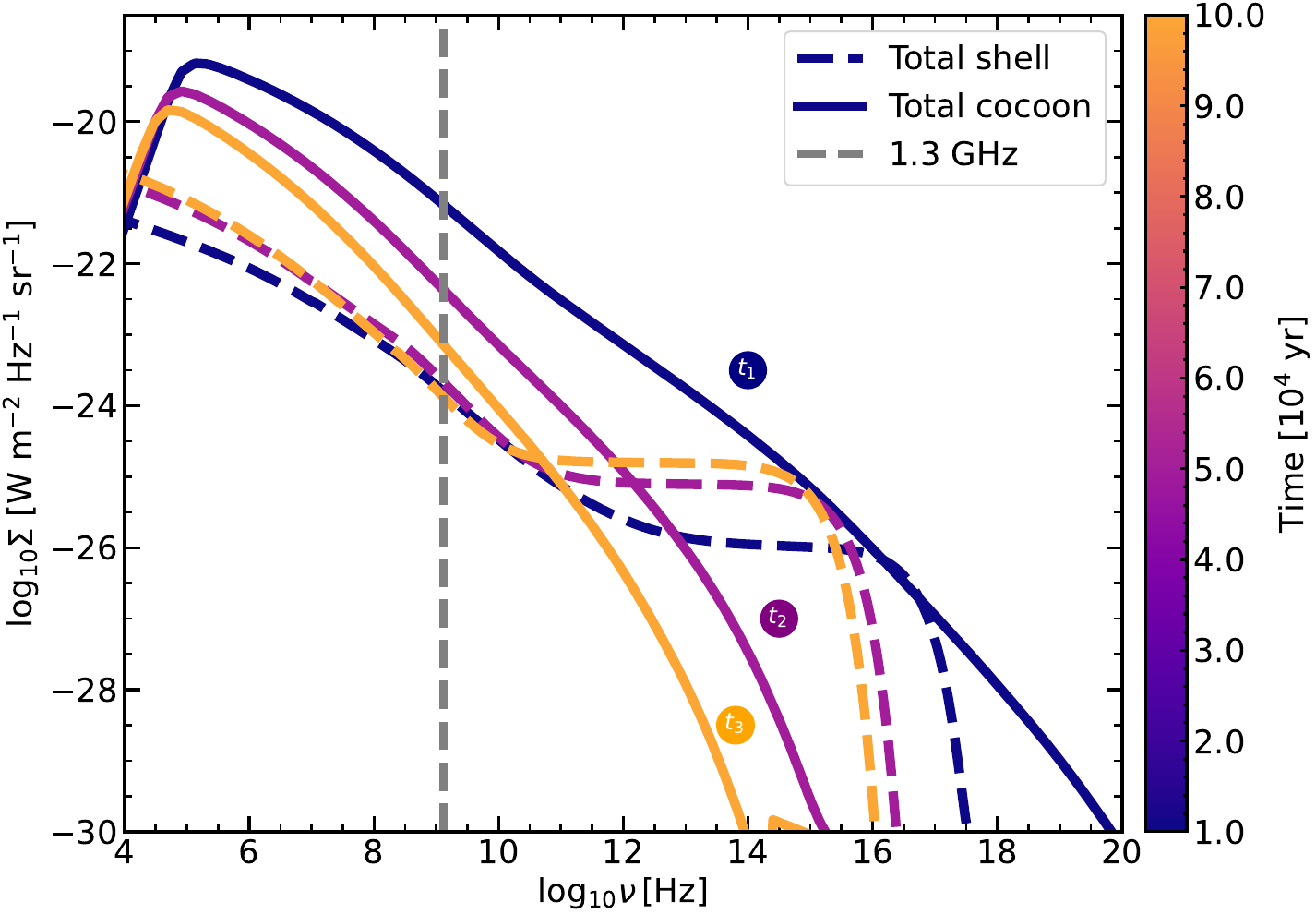}
  \caption{Surface brightness of the cocoon (solid lines) and shell (dashed lines) for the super-Eddington (top) and sub-Eddington (bottom) systems, again for three different epochs of the MQR life ($t_{1,2,3}$).}
  \label{fig: surfacebright}
\end{figure}

\hyperref[fig: surfacebright]{Figure~\ref{fig: surfacebright}} shows the surface brightness of the MQRs at a fiducial Galactic distance of $d=2\,$kpc, for the two scenarios considered, and for each of the three selected times considered for the SEDs. We show the total emission of the shell (dashed lines) and cocoon (solid lines) separately, because their emitting surfaces differ and we aim to distinguish their contributions. We also show in gray the frequency $1.3\,{\rm GHz}$, which is of interest for radio detectability. 

For reference, we also estimate the local effective radio spectral index around \(1.3\,\mathrm{GHz}\). From the modeled surface-brightness spectra of the cocoon, we obtain at the initial epoch \(\alpha_{\rm r, super}\simeq 0.67\) for the super-Eddington case and \(\alpha_{\rm r,sub}\simeq 0.83\) for the sub-Eddington case (where \(\Sigma_\nu \propto \nu^{-\alpha_{\rm r}}\)). 

\subsection{Synthetic maps}\label{sec: synthetic}

\begin{figure}
  \centering
  \includegraphics[
  width=0.95\columnwidth]
{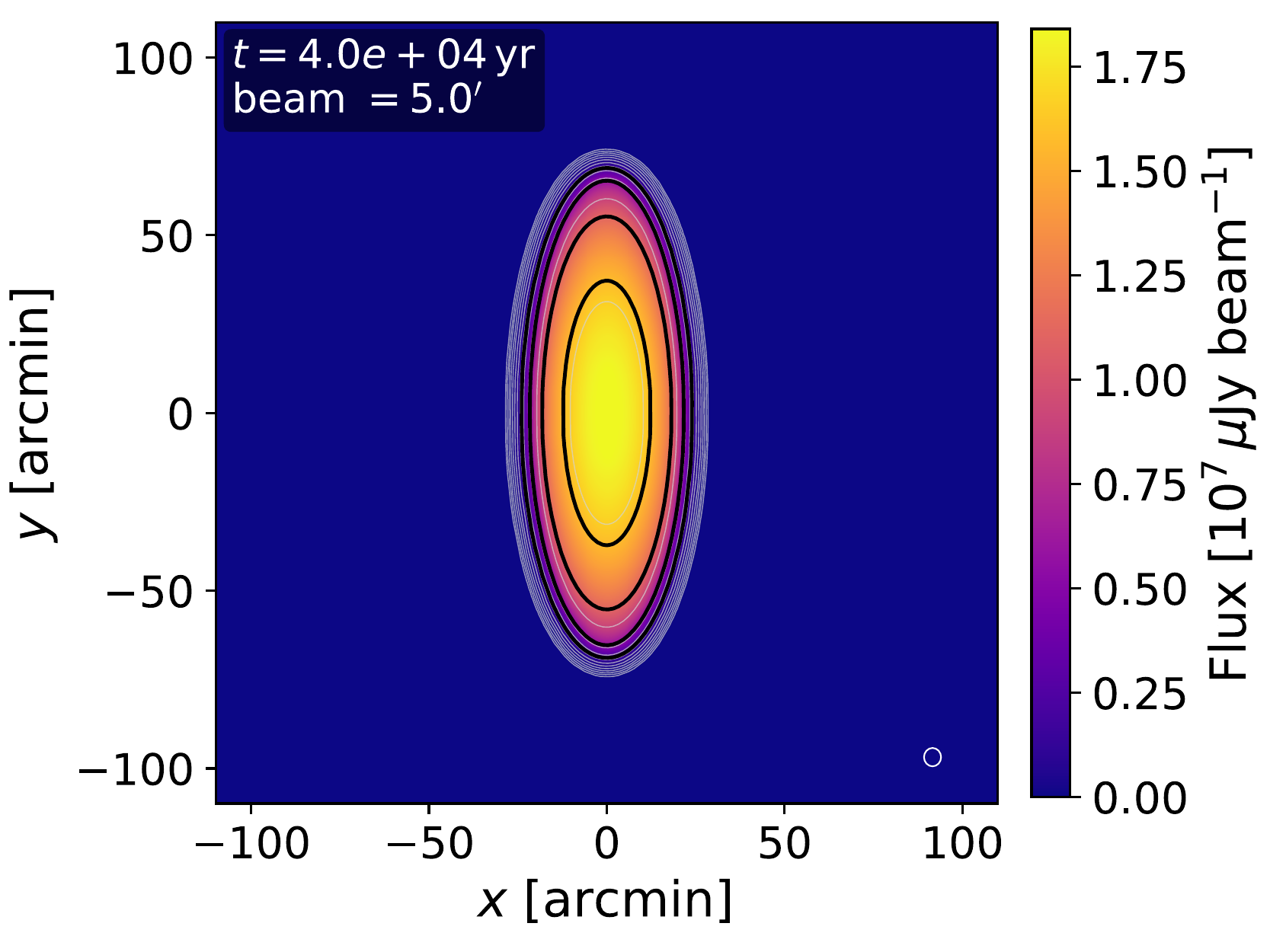}
  \includegraphics[
  width=0.95\columnwidth]{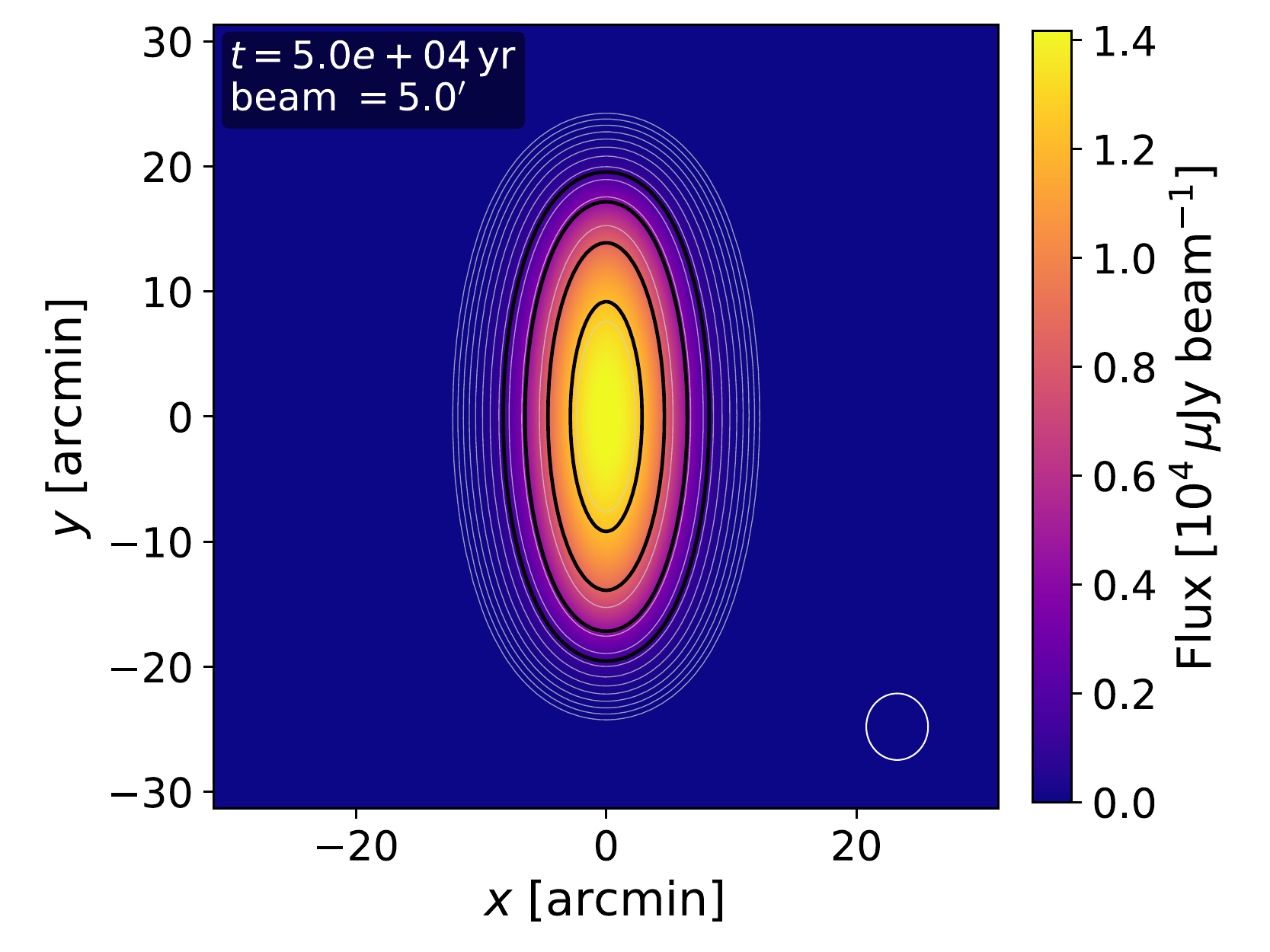}
  \caption{Synthetic radio maps for the super- (top) and sub-Eddington (bottom) scenarios, for an intermediate epoch (40 and 50 kyr, respectively). The white circle in the lower corner indicates the 5'
 FWHM Gaussian beam. The instrument considered is the radio interferometer MeerKAT-like. We notice that the two panels use different color scales.}
  \label{fig: synthetic_radio}
\end{figure}

\hyperref[fig: synthetic_radio]{Figure~\ref{fig: synthetic_radio}} shows the synthetic radio maps at $1.3\,{\rm GHz}$ for the super- and sub-Eddington scenarios. The maps were obtained by projecting the MQR onto the plane of the sky, assuming an inclination angle of $90^\circ$, and by smoothing the resulting intensity distribution with a Gaussian beam of $5'$ FWHM. The source is centered at $(x,y)=(0,0)$, and the axes are given in arcmin. The color scale shows the flux density per beam, in units of $\mu{\rm Jy\,beam^{-1}}$. No diffuse radio background is included in these maps. 

In both radio and X-ray maps, the contours represent source-only isophotes. The black contours correspond to fixed fractions of the peak value in each panel, whereas the lighter contours show additional logarithmically spaced levels to emphasize the projected morphology. The color scales are chosen independently for the super- and sub-Eddington cases.

The radio emission has a smooth and elongated morphology, reflecting the assumed prolate geometry of the cocoon. In the super-Eddington case, the source extends over angular scales of the order of one degree at the epoch shown. As indicated by the temporal evolution of the 1.3 GHz surface brightness in \hyperref[fig: surfacebright]{Fig.~\ref{fig: surfacebright}}, the peak radio brightness decreases as the system expands. This decrease is caused by the combined effect of particle cooling, magnetic-field dilution, and the increase of the projected emitting area. In the sub-Eddington case, the same qualitative morphology is obtained, but the radio brightness is several orders of magnitude lower. This illustrates that, even if the integrated non-thermal luminosity is non-negligible, direct radio detection of low-power MQRs is expected to be strongly limited by surface brightness and by the recovery of large angular scales.

%MAPAS RADIO 

\begin{figure}[ht!]
  \centering
  \includegraphics[
  width=0.95\columnwidth]
{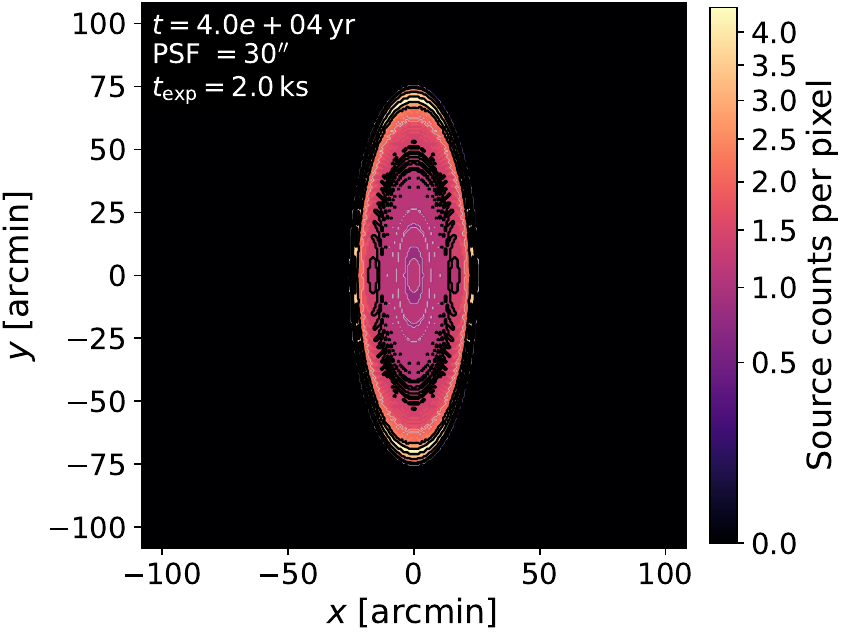}
  \includegraphics[
  width=0.95\columnwidth]{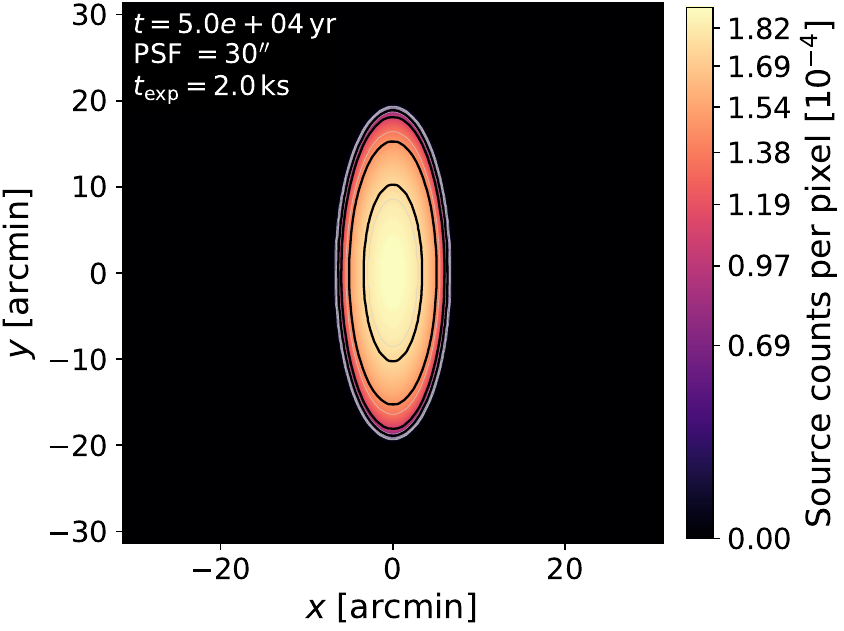}

  \caption{Synthetic X-ray maps for the super- (top) and sub-Eddington (bottom) scenarios, for an intermediate epoch (40 and 50 kyr, respectively). The instrument considered is the \textit{eROSITA}-like X-ray telescope. We notice that the two panels use different color scales.}
  \label{fig: synthetic_RX}
\end{figure}

\hyperref[fig: synthetic_RX]{Figure~\ref{fig: synthetic_RX}}  shows the corresponding synthetic soft X-ray maps for the super- and sub-Eddington scenarios. We integrate the emission in the $0.5$--$2\,{\rm keV}$ band and convert the intrinsic surface brightness into expected source counts per pixel using an \textit{eROSITA}-like exposure of $2\,{\rm ks}$ and a Gaussian PSF with FWHM $=30''$. The maps show source counts only; instrumental background, diffuse Galactic emission, spatially varying exposure, and detailed absorption effects are not included.

The X-ray morphology differs from the radio morphology because the soft X-ray emission is dominated by the shocked shell. In the super-Eddington scenario, the shell could be detectable under favorable foreground and exposure conditions. As the remnant expands and decelerates, the post-shock temperature decreases and the emission in the $0.5$--$2\,{\rm keV}$ band fades, leaving a progressively fainter shell-like structure. In the sub-Eddington scenario, the expected number of counts per pixel is very low for the adopted exposure, indicating that such systems would be difficult to detect with shallow soft-X-ray observations even under the optimistic assumptions adopted here. Therefore, the synthetic maps reinforce the conclusion drawn from the surface-brightness estimates: powerful MQRs may be detectable as extended radio and soft-X-ray structures, whereas low-power remnants, arguably the most common ones, are likely to require favorable environments, low foreground contamination, or deeper observations.

The MeerKAT Galactic Plane Survey reaches a typical rms sensitivity of \(\sim10{-}20\,\mu{\rm Jy\,beam^{-1}}\) at \(\sim8''\) resolution \citep{Meerkat2024}, but the detectability of degree-scale diffuse emission is mainly limited by large-scale flux recovery, calibration systematics, source confusion, and the Galactic radio background. Similarly, eROSITA point-source flux limits in the \(0.5{-}2\) keV band are of order \(10^{-14}{-}10^{-13}\,{\rm erg\,s^{-1}\,cm^{-2}}\), but extended sources in the Galactic plane are more severely affected by background and by photoelectric absorption below \(\sim1\) keV \citep{eRositaBook2012}. The synthetic maps should therefore be interpreted as optimistic, source-only estimates rather than as full detectability simulations.

\subsection{Scan of parameter space}\label{sec: scan}

We explored the parameter space, assuming the same jet powers for the super- and sub-Eddington systems $(10^{40};10^{37}\,{\rm erg\,s^{-1}})$, and vary the ISM density in which the MQR evolves $(10^{-2};1\,{\rm cm^{-3}})$ and the diffusion regime (Kolmogorov, 1/3; or Kraichnan, 1/2). 

\begin{figure}  
  \centering
    \centering
    \includegraphics[width=\columnwidth]{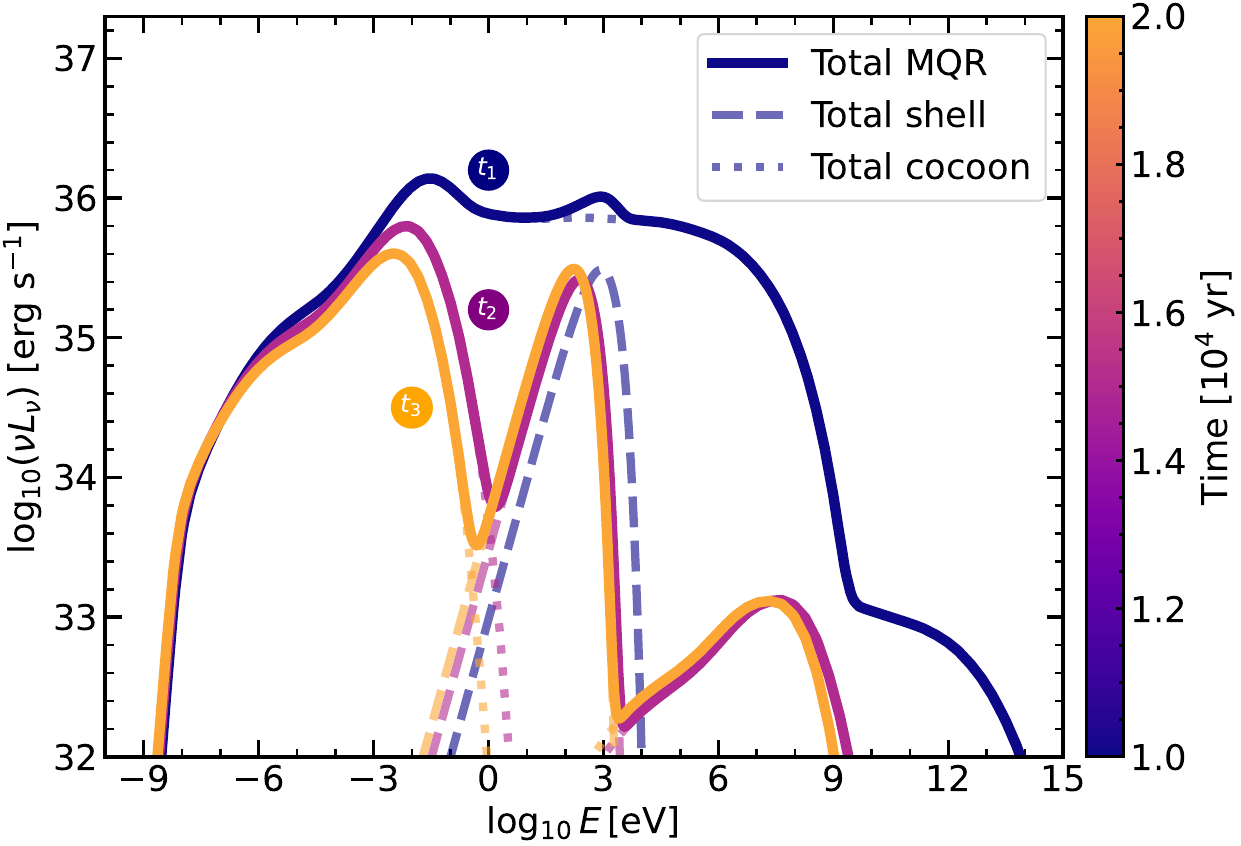}
    \centering
    \includegraphics[width=\columnwidth]{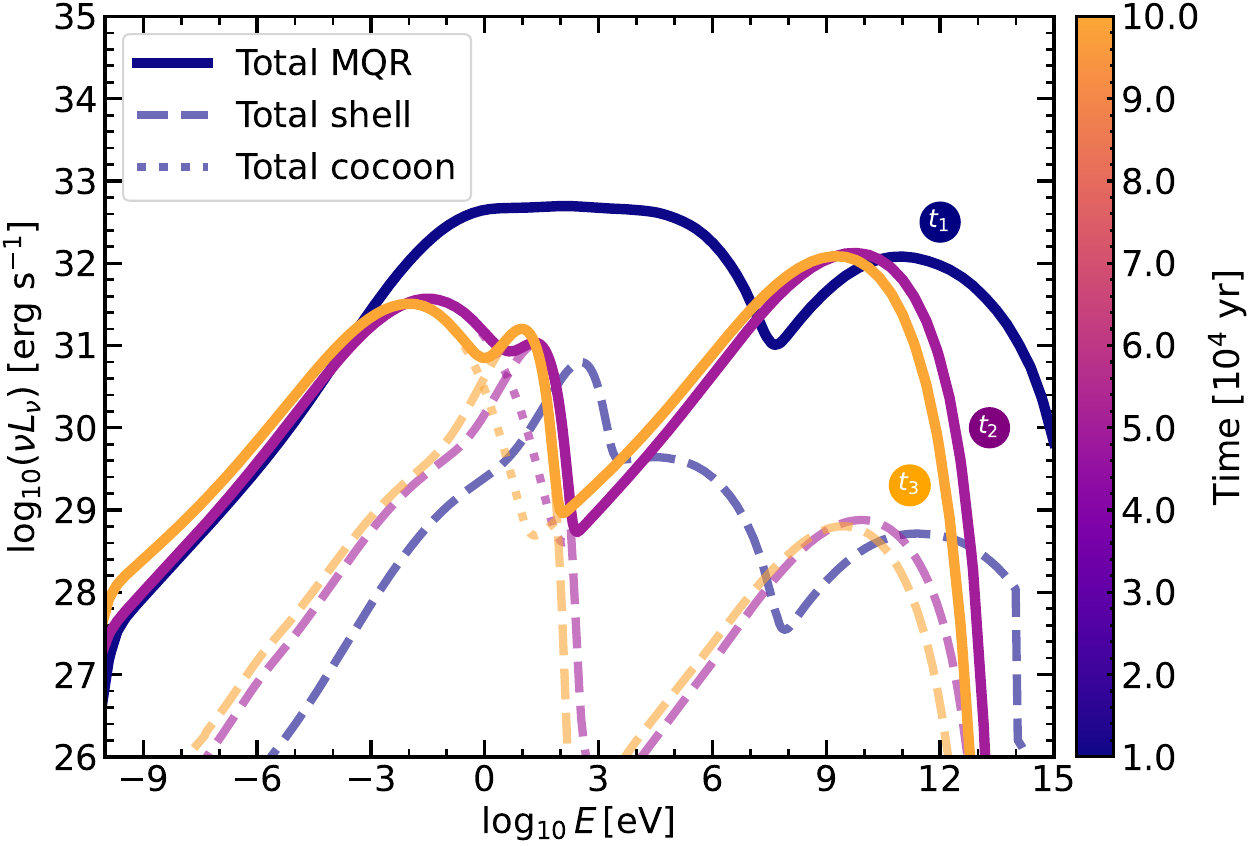}

  \caption{Same as \hyperref[fig: SED_total]{Fig.~\ref{fig: SED_total}}, but for the two extreme alternative scenarios explored in \hyperref[sec: scan]{Sect.~\ref{sec: scan}}. The solid curves correspond to the total MQR emission, i.e. the sum of the shell and cocoon components.}
  \label{fig: app_SED_total}
\end{figure}

Therefore, we have eight alternative scenarios. For each of them, we performed the same calculations as for the fiducial models. We only show here the SEDs for the extreme scenarios: the sub-Eddington system with a low ambient density $(L_{\rm j}=10^{37}\,{\rm erg\,s^{-1}};n_{\rm ISM}=10^{-2}\, \,{\rm cm^{-3}})$ and the super-Eddington system with a high ambient density $(L_{\rm j}=10^{40}\,{\rm erg\,s^{-1}};n_{\rm ISM}=1\, \,{\rm cm^{-3}})$, both for a Kraichnan-like diffusion regime. \hyperref[fig: app_SED_total]{Figure~\ref{fig: app_SED_total}} shows the total SEDs for three different epochs of the MQRs, for both scenarios.

The full set of eight parameter-scan models shows that the jet power is the main parameter controlling the overall normalization of the SED. Sub-Eddington remnants reach peak values of \(\nu L_\nu \sim 10^{32}-10^{34}\,{\rm erg\,s^{-1}}\), whereas super-Eddington remnants can reach \(\nu L_\nu \sim 10^{36}-10^{37}\,{\rm erg\,s^{-1}}\). This contrast is broadly consistent with the difference in injected mechanical power.

Variations in the ambient density mainly change the relative importance of the shell and cocoon components. Denser environments produce more efficient shell emission because the swept-up material provides a larger target density for thermal Bremsstrahlung, non-thermal Bremsstrahlung, and hadronic interactions. Conversely, low-density environments lead to more diluted remnants and weaker shell-related emission, so that the cocoon contribution becomes relatively more important. The diffusion prescription has a more limited impact on the bolometric appearance of the SEDs, but it modifies the high-energy tails by changing the escape efficiency of the most energetic particles. In particular, moving from Kolmogorov-like to Kraichnan-like diffusion mainly affects the high-energy gamma-ray output and the detailed shell/cocoon balance at late times.

Overall, the parameter scan indicates that the qualitative conclusions derived from the fiducial models are robust: powerful MQRs can remain bright enough to produce detectable intrinsic counterparts under favorable conditions, whereas low-power remnants are much fainter and their direct identification is expected to be limited primarily by surface brightness.

\section{Discussion}\label{sec:discussion}

The results presented above show that the direct identification of MQRs is primarily a surface-brightness problem. Although the total non-thermal luminosity of the cocoon can remain significant during a substantial fraction of the remnant lifetime, the emission is distributed over angular scales of several tens of arcminutes to degrees for nearby Galactic sources. As a consequence, MQRs can be relevant in an integrated energetic sense while remaining difficult to detect in conventional surveys. This effect is particularly important in the radio band, where the synchrotron emission from the cocoon is expected to be smooth and extended, and where interferometric observations may filter out a large fraction of the flux if the shortest spatial frequencies are not recovered \citep{Thompson2017}.

The presence of radio emission in the cocoon does not require a magnetic field much larger than that of the ambient ISM. Even when the cocoon magnetic field approaches the imposed ISM floor at late times, relativistic electrons confined within the remnant still radiate through synchrotron emission. The limiting factors are, instead, the dilution of the particle population and the large projected area of the source. Electrons escaping into the surrounding ISM would produce an even more diluted component, unless they encounter regions of enhanced magnetic field or dense material. Therefore, the expected radio counterpart of an MQR is faint, smooth, and sensitive to the recovery of large angular scales. This naturally explains why extended MQ bubbles can be difficult to identify despite the large kinetic energy injected by the jets.

The synthetic MeerKAT-like maps illustrate this point. In the super-Eddington scenario, the remnant can produce an extended radio cocoon potentially detectable as a low-surface-brightness structure if the large-scale emission is recovered. In the sub-Eddington scenario, the predicted radio emission is much fainter, although its detectability may improve in regions where the Galactic diffuse background and source confusion are lower. For this reason, MQRs located away from the Galactic plane may be more favorable targets for direct radio searches, even if the probability of interaction with dense molecular material is smaller. Conversely, MQRs in the plane may be more likely to illuminate nearby clouds in gamma rays, but their intrinsic radio cocoon could remain hidden by the diffuse Galactic foreground.

The soft X-ray band provides a complementary diagnostic. In our model, the X-ray emission is dominated by the shocked shell, especially in young and powerful systems. The thermal component traces the expansion of the cocoon into the ISM and therefore probes the hydrodynamical impact of the former jet activity. However, the X-ray detectability is strongly time dependent. As the shell expands and decelerates, the post-shock temperature decreases and the emission progressively shifts to lower energies. Thus, even if the total thermal luminosity remains relevant, the fraction emitted in the 0.5--2 keV band can decrease at late times. The \textit{eROSITA}-like maps should therefore be interpreted as first-order estimates of the expected count distribution. A detailed comparison with observations would require the full instrumental response, realistic absorption, spatially varying exposure, and background modeling.

We notice that, although the cocoon keeps expanding during the remnant phase, its angular growth is expected to be negligible on observational timescales. For the representative intermediate epochs considered here, \(t=4\times10^{4}\,\mathrm{yr}\) for the super-Eddington model and \(t=5\times10^{4}\,\mathrm{yr}\) for the sub-Eddington model, the expansion law \(l_{\rm c}\propto t^{2/5}\) implies \(v_{\rm c}=(2/5)l_{\rm c}/t\). The corresponding growth of the semi-major axis is only \(\sim 4\times10^{-4}\,\mathrm{pc\,yr^{-1}}\) and \(\sim 9\times10^{-5}\,\mathrm{pc\,yr^{-1}}\), respectively. At a fiducial distance of \(2\,\mathrm{kpc}\), this translates into angular changes of \(\sim 0.04''\,\mathrm{yr^{-1}}\) and \(\sim 0.01''\,\mathrm{yr^{-1}}\), or only \(\sim 0.4''\) and \(\sim 0.1''\) over a decade. Therefore, proper expansion is not expected to provide a practical diagnostic; MQR identification should instead rely on their broadband spectrum, large-scale morphology, surface brightness, and environmental context.

Stochastic re-acceleration can play an important role in shaping the non-thermal emission ---mainly regarding proton interactions--- depending on the MQ power and diffusion regime. In the present model, second-order Fermi acceleration does not replace the first-order acceleration that occurred at the jet termination region during the active MQ phase. Instead, it redistributes particle energy inside the turbulent cocoon after jet shutdown. This process can partially compensate for adiabatic and radiative losses of the particles. The resulting emission depends not only on the original jet power, but also on the level of turbulence, its decay timescale, and the diffusion regime inside the cocoon. For protons, radiative losses inside the dilute cocoon are inefficient; therefore, turbulent confinement and re-acceleration mainly affect their residence time and the energy spectrum of the particles eventually escaping into the environment. This is particularly relevant for cloud-illumination scenarios, in which long-lived hadronic reservoirs can produce gamma rays after the central engine has become inactive.

These properties provide possible criteria for distinguishing MQRs from other extended Galactic sources. Supernova remnants can also show shell-like radio and X-ray morphologies, but their energy is deposited impulsively by an explosion, whereas MQRs are inflated by collimated jets over the active lifetime of the binary. As a result, MQRs are expected to be more elongated or bipolar, especially in relatively homogeneous environments. Pulsar wind nebulae and TeV halos, on the other hand, are usually centered on an energetic pulsar and are continuously powered by its spin-down luminosity. In contrast, an MQR may lack any currently active compact counterpart. Moreover, in hadronic scenarios the gamma-ray emission can be spatially displaced from the MQR toward nearby dense clouds illuminated by escaping protons. The combination of an extended radio/X-ray cocoon, weak or absent central activity, and offset gamma-ray emission would therefore provide a strong indication of an MQR.

The comparison with unidentified gamma-ray sources in the super-Eddington scenarios is particularly relevant. Instruments such as LHAASO and  the HAWC are well suited to detect extended UHE gamma-ray emission, but their angular resolution and limited multiwavelength information can make it difficult to identify the accelerator. The Cherenkov Telescope Array (CTA) will provide better angular resolution and spectral information in the TeV range, helping to distinguish between emission associated with the cocoon and emission arising from nearby illuminated clouds. At lower energies, \textit{Fermi}-LAT can constrain the GeV counterpart, although confusion with diffuse Galactic emission becomes important for very extended sources. A multi-instrument strategy is therefore required: radio and soft X-rays can reveal the remnant structure, while gamma rays trace the highest-energy particles and their interaction with the surrounding medium.

The discussion of extended-source detectability by \cite{Celli&Peron2024} is directly relevant in this context. These authors show that gamma-ray detection prospects for extended sources should be evaluated using extension-dependent sensitivities, rather than point-source sensitivities alone. They also emphasize that improved spectral and morphological information
from next-generation Imaging Atmospheric Cherenkov Telescopes (IACTs) will be needed to clarify the nature of many extended UHE sources detected by LHAASO.

Several limitations of the present model should be kept in mind. We assume a smooth, homogeneous cocoon and a uniform ambient ISM, whereas real MQ environments can be highly structured. Density gradients and anisotropic turbulence, for instance, can modify both the morphology and the radiative output of the remnant. The magnetic field and turbulence evolution are described phenomenologically, and the stochastic acceleration efficiency is constrained through an energetic cap rather than derived from magnetohydrodynamical simulations. In addition, the synthetic radio maps do not include a full interferometric response, and the \textit{eROSITA}-like maps neglect the detailed detector response, background components, and line-of-sight absorption. The results should therefore be regarded as physically motivated detectability estimates rather than predictions for a specific observation.

Our treatment of the shell assumes a smooth, hot, and adiabatic post-shock region whose pressure adjusts to that of the cocoon. We therefore do not model internal substructure or radiative fragmentation of the swept-up material. In a more realistic, inhomogeneous medium, denser regions of the shell could cool earlier than the average flow, increasing the target density for $pp$ interactions and Bremsstrahlung emission, enhancing the thermal component, and promoting instabilities that may disrupt the large-scale shell structure at earlier times. A detailed hydrodynamical treatment of this radiative and clumpy stage is beyond the scope of the present work, but it will be necessary to refine the observational predictions for evolved MQRs.

Finally, a zeroth-order estimate of the Galactic population of MQRs can be obtained by assuming a quasi-steady state between the active MQ phase and the remnant phase. Galactic X-ray binary catalogs list \(\sim 5\times10^{2}\) known X-ray binaries \citep{Liu_etal_2006_catalogoHMXRB,Fortin_etal_2023_catologoHMXRB,Neumann_etal_2023_catalogoHMXRB}, although only a small subset of them are expected to launch relativistic jets powerful and long-lived enough to inflate large-scale cocoons . If \(N_{\rm MQ}\) is the number of such active systems in the Galaxy, the expected number of remnants is
\[
N_{\rm MQR} \sim N_{\rm MQ}\frac{t_{\rm MQR}}{t_{\rm MQ}} .
\]
For the fiducial models adopted here, \(t_{\rm MQ}=10^{4}\,\mathrm{yr}\), whereas \(t_{\rm MQR}\sim7.5\times10^{4}\,\mathrm{yr}\) and \(\sim3\times10^{5}\,\mathrm{yr}\) for the super- and sub-Eddington cases, respectively. This gives \(N_{\rm MQR}\sim7.5\,N_{\rm MQ}\) and \(\sim30\,N_{\rm MQ}\). Thus, even if only a few to a few tens of Galactic X-ray binaries undergo a suitable MQ phase, the Milky Way could host from several tens to a few hundred MQRs. This number should be regarded as an upper limit to the detectable population, since not all MQs are expected to produce sufficiently powerful or long-lived jets, and the radio or X-ray surface brightness of the remnant may fall below observational thresholds before the hydrodynamical end time.

\section{Conclusions}\label{sec:conclusions}

We explored the intrinsic emission and detectability of MQRs, focusing on the long-lived cocoon and its shocked shell after jet activity has ceased. We considered both super- and sub-Eddington systems and followed the time-dependent evolution of relativistic particles, including stochastic re-acceleration in the turbulent cocoon. Our results show that MQRs can remain energetically relevant after jet shutdown, but their direct identification is mainly limited by surface brightness rather than by total luminosity.

In the radio band, the emission is dominated by synchrotron radiation from electrons confined in the cocoon, producing a smooth, extended, and elongated morphology. Powerful MQRs may be detectable as diffuse radio structures if large angular scales are recovered, whereas sub-Eddington remnants are expected to be much harder to identify directly. In soft X-rays, the dominant contribution comes from thermal Bremsstrahlung in the shocked shell, which is strongest in young and powerful systems but fades as the shell expands and decelerates.

The synthetic MeerKAT-like and \textit{eROSITA}-like maps reinforce this picture. Super-Eddington remnants can produce extended radio and soft-X-ray counterparts under favorable conditions, while low-power remnants require deeper observations, lower foreground contamination, or particularly favorable environments. These results support the idea that MQRs may constitute a hidden population of Galactic non-thermal sources, whose identification will require multiwavelength searches combining radio, X-ray, and gamma-ray information.

\begin{acknowledgements}
We thank the referee for his/her comments. LA thanks the Universidad Nacional de La Plata. GER and VB-R were funded by PID2022-136828NB-C41/AEI/10.13039/501100011033/ and PID2025-168247NB-C41/AEI/10.13039/501100011033/, and through the ``Unit of Excellence María de Maeztu'' award to the Institute of Cosmos Sciences (CEX2019-000918-M, CEX2024-001451-M). VB-R is Correspondent Researcher of CONICET, Argentina, at the IAR.
\end{acknowledgements}

\bibliographystyle{aa} 
\bibliography{main}

\begin{appendix}

\section{Characteristic end times of the remnant}
\label{app:end_times}

We distinguish between two characteristic timescales associated with the late evolution of the shell. The first one is the radiative transition time, $t_{\rm rad}$, defined as the time at which the post-shock cooling time becomes comparable to the dynamical time of the expanding shell. For a strong shock, the post-shock temperature is estimated as in Eq. \eqref{eq:Tps}, while the cooling time is approximated by
\begin{equation}
    t_{\rm cool} =
    \frac{3 k_{\rm B}T_{\rm sh}}
    {n_{\rm ps}\Lambda(T_{\rm sh})},
\end{equation}
where $n_{\rm sh}=4n_{\rm ISM}$ for a strong shock and $\Lambda(T)$ is the adopted radiative cooling function (see, e.g., \citealt{Abaroa_etal_2023}). We define $t_{\rm rad}$ through
\begin{equation}
    t_{\rm cool}(t_{\rm rad}) = t_{\rm dyn}(t_{\rm rad}),
    \qquad
    t_{\rm dyn} = \frac{R_{\rm sh}}{v_{\rm c}} .
\end{equation}
This criterion marks the time at which the adiabatic-shell approximation ceases to be appropriate because radiative losses become dynamically important.

However, for low-power systems the shell can become dynamically weak before reaching the formal radiative transition. We therefore also estimate a dynamical mixing time, $t_{\rm mix}$, defined as the time at which the shell expansion velocity becomes comparable to the effective propagation speed of disturbances in the ambient medium,
\begin{equation}
    v_{\rm c}(t_{\rm mix}) = c_{\rm eff}.
\end{equation}
We take
\begin{equation}
    c_{\rm eff}^2 = c_{\rm s}^2 + v_{\rm A}^2 + \sigma_{\rm turb}^2 ,
\end{equation}
where $c_{\rm s}$ is the sound speed of the ambient gas, $v_{\rm A}$ is the Alfv\'en speed, and $\sigma_{\rm turb}$ is an optional turbulent velocity dispersion. This criterion does not describe radiative cooling, but rather the stage at which the remnant is no longer expected to remain a strong, coherent, supersonically expanding shell.

After jet shutdown, the shell evolution is approximated as an impulsive expansion. Thus, for a given value of $c_{\rm eff}$, the mixing time can be estimated from
\begin{equation}
    t_{\rm mix} =
    t_0
    \left[
    \frac{v_{\rm c}(t_0)}{c_{\rm eff}}
    \right]^{5/3}.
\end{equation}

We calculate both $t_{\rm rad}$ and $t_{\rm mix}$, and define the characteristic hydrodynamical end time of the remnant as
\begin{equation}
    t_{\rm end} = \min(t_{\rm rad}, t_{\rm mix}) .
\end{equation}
The limiting process is therefore identified as either radiative cooling or dynamical mixing, depending on which condition is reached first. This distinction is particularly relevant for low-power jets. For example, for the scenario with $L_{\rm j}=10^{37}\,{\rm erg\,s^{-1}}$, $t_{\rm 0}=10^4\,{\rm yr}$, $n_{\rm ISM}=0.1\,{\rm cm^{-3}}$, and $B_{\rm ISM}=3\,\mu{\rm G}$, the formal radiative transition can occur at $t_{\rm rad}\sim 10^6\,{\rm yr}$, whereas the shell velocity becomes comparable to a typical warm magnetized ISM effective speed, $c_{\rm eff}\sim 20$--$30\,{\rm km\,s^{-1}}$, at $t_{\rm mix}\gtrsim 10^5\,{\rm yr}$. In such cases, the shell may dynamically merge with the ambient medium before becoming radiative.

Finally, we emphasize that $t_{\rm end}$ is a hydrodynamical timescale and is distinct from the observational lifetime of the remnant. In some models, the predicted radio or X-ray surface brightness becomes negligible well before either $t_{\rm rad}$ or $t_{\rm mix}$. Therefore, we also track when the source falls below the relevant observational thresholds, and we do not necessarily evolve all low-power models up to the formal radiative transition time if the emission has already become observationally insignificant.

\section{Effect of diffusion and stochastic re-acceleration}\label{appendix}

\begin{figure*}  
  \centering    \includegraphics[width=\linewidth]{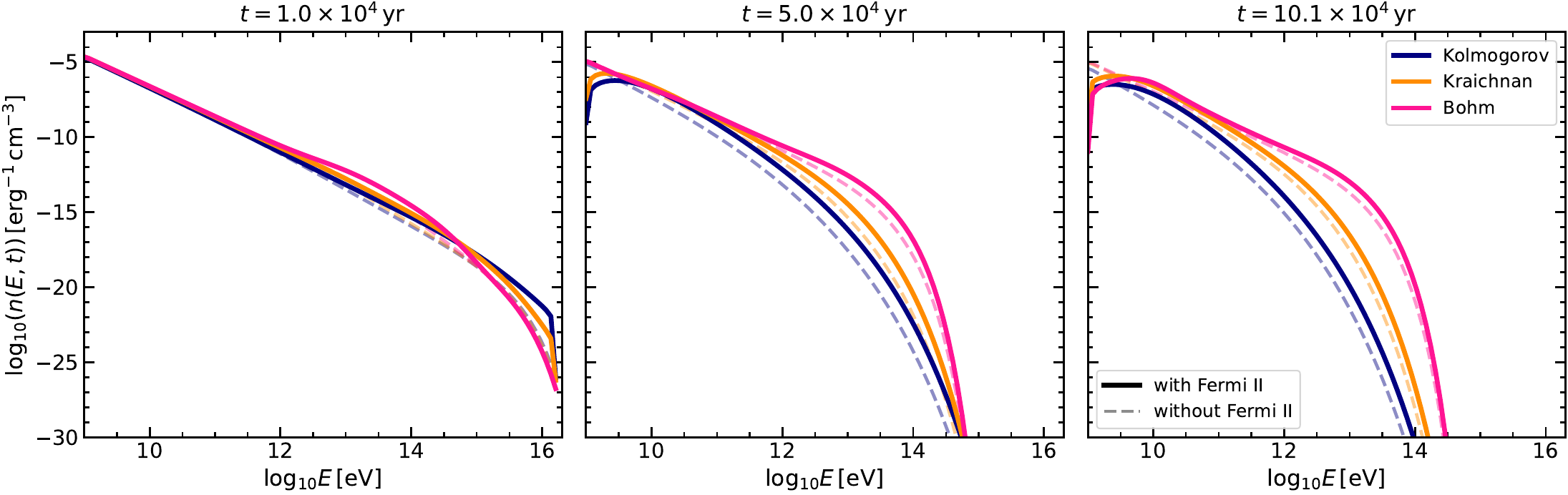}
  \caption{Proton distributions in the cocoon for the sub-Eddington scenario of our fiducial model from the main text. We show panels for three different epochs of the MQR: initial (left), intermediate (middle), and late (right) epochs. In each panel we show the proton distribution for three different diffusion regimes (Kolmogorov, Kraichnan, and Bohm), with and without re-acceleration (solid and dashed lines, respectively).}
  \label{fig: app_distributionsSub}
\end{figure*}

\begin{figure*}  
  \centering    \includegraphics[width=\linewidth]{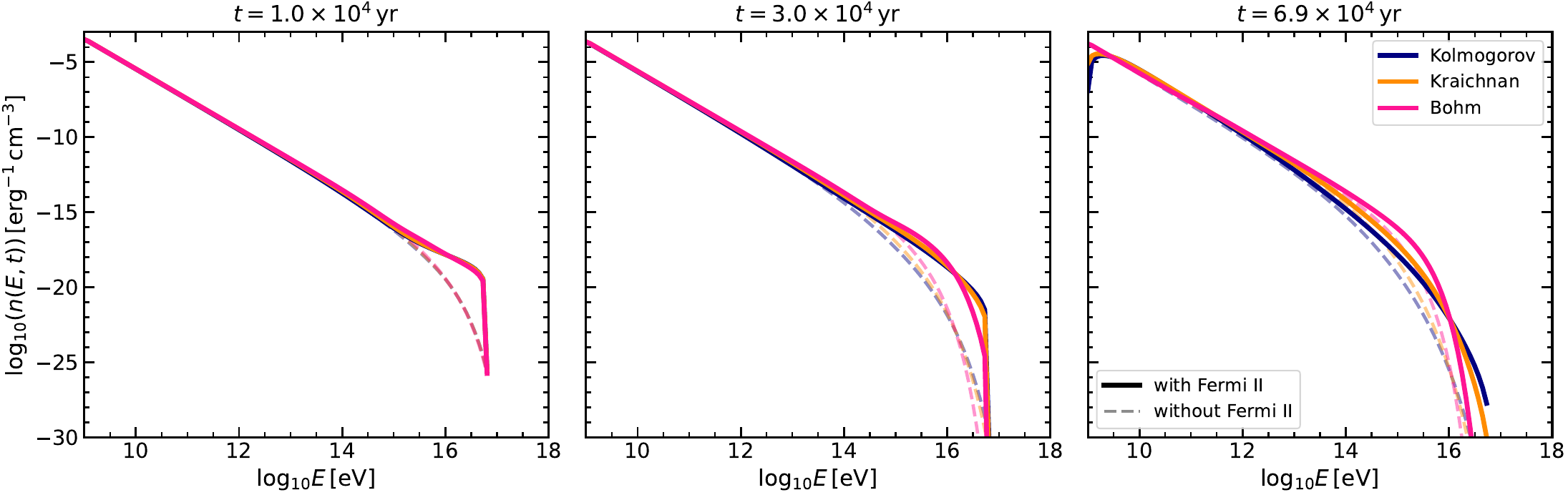}
  \caption{Same as in \hyperref[fig: app_distributionsSub]{Fig.~\ref{fig: app_distributionsSub}} but for the super-Eddington scenario of our fiducial model from the main text.}
  \label{fig: app_distributionsSuper}
\end{figure*}

\hyperref[fig: app_distributionsSub]{Figures~\ref{fig: app_distributionsSub}} and \hyperref[fig: app_distributionsSub]{\ref{fig: app_distributionsSuper}} isolate the effect of the diffusion prescription and stochastic re-acceleration on the proton population in the fiducial sub- and super-Eddington models. Since radiative losses are inefficient for protons in the dilute cocoon, the main differences arise from confinement and energy-space diffusion. 

Bohm-like transport produces the strongest confinement, whereas Kolmogorov and Kraichnan diffusion lead to faster escape at high energies. The comparison with the runs without Fermi-II acceleration shows that stochastic re-acceleration mainly affects the high-energy tail of the retained proton population, and therefore the spectrum of particles that can later escape and illuminate nearby clouds.

\section{Reverse shock formation in the shell}\label{appendix_shock}

We can estimate whether a reverse shock is expected to form in the shell by comparing two timescales: the time required for pressure perturbations to cross the shocked shell, and the time over which the shell velocity changes appreciably.

\subsection{Changes in velocity}
In the remnant phase, the cocoon semi-major axis is assumed to evolve with an expansion velocity given by
\begin{equation}
    v_{\rm c}(t) = \frac{{\rm d} l_{\rm c}}{{\rm d}t}
    = \frac{2}{5}\frac{l_{\rm c}(t)}{t}.
\end{equation}
Since $v_{\rm c}\propto t^{-3/5}$, its time derivative is
\begin{equation}
    \dot{v}_{\rm c}(t)
    = -\frac{3}{5}\frac{v_{\rm c}(t)}{t}.
\end{equation}
Therefore, the characteristic deceleration time of the shell is
\begin{equation}
    t_{\rm dec}
    \equiv \left|\frac{v_{\rm c}}{\dot{v}_{\rm c}}\right|
    = \frac{5}{3}t.
\end{equation}
This is the timescale over which the expansion velocity changes significantly.

\subsection{Changes in pressure}

The sound speed in the post-shock gas of the shell is given by
\begin{equation}
    c_{\rm s,sh}
    =
    \left(\frac{\Gamma_{\rm sh} k_{\rm B} T_{\rm sh}}{\mu m_{\rm p}}\right)^{1/2},
\end{equation}
where $\Gamma_{\rm sh}$ is the adiabatic index of the shocked shell. Substituting the expression for $T_{\rm ps}$ gives
\begin{equation}
    c_{\rm s,ps}
    =
    \left(\frac{3\Gamma_{\rm sh}}{16}\right)^{1/2}v_{\rm c}.
\end{equation}
For a non-relativistic monoatomic plasma, $\Gamma_{\rm sh}=5/3$, and therefore
\begin{equation}
    c_{\rm s,ps}
    =
    \frac{\sqrt{5}}{4}v_{\rm c}
    \simeq 0.56\,v_{\rm c}.
\end{equation}

The shell thickness follows from mass conservation of the swept-up ISM and the strong-shock compression ratio. The sound-crossing time across the shell is therefore
\begin{equation}
    t_{\rm sc,sh}
    \equiv
    \frac{\Delta l_{\rm c}}{c_{\rm s,ps}}
    \simeq
    \frac{0.1\,l_{\rm c}}{\left(3\Gamma_{\rm sh}/16\right)^{1/2}v_{\rm c}}.
\end{equation}
Using \(v_c=(2/5)l_c/t\), this becomes
\[
t_{\rm sc,sh}=\frac{t}{\sqrt{3\Gamma_{\rm sh}}}.
\]
For \(\Gamma_{\rm sh}=5/3\),
\[
t_{\rm sc,sh}=\frac{t}{\sqrt{5}}\simeq 0.45\,t.
\]
The relevant ratio is therefore
\[
\frac{t_{\rm sc,sh}}{t_{\rm dec}}
=
\frac{3}{5\sqrt{3\Gamma_{\rm sh}}}
=
\frac{3}{5\sqrt{5}}
\simeq 0.27.
\]
This estimate indicates that the shocked shell can communicate pressure changes across its thickness on a timescale shorter than the deceleration time. Therefore, within the smooth one-zone description adopted here, the shell dynamics can be accounted for by the outer forward shock alone, and a persistent reverse shock inside the shell is not expected to form as a generic feature.

\section{Synthetic surface-brightness maps}\label{app: maps}

We model the emitting region as a prolate cocoon surrounded by a prolate 
shell. 

For a given inclination angle, the Cartesian coordinates are first 
rotated with respect to the line of sight. We then define two 
ellipsoidal radii,
\[
R_{\rm in}(x,y,z;t)=
\left[
\left(\frac{x'}{l_c(t)}\right)^2+
\left(\frac{y'}{w_c(t)}\right)^2+
\left(\frac{z'}{w_c(t)}\right)^2
\right]^{1/2},
\]
and
\[
R_{\rm out}(x,y,z;t)=
\left[
\left(\frac{x'}{l_{\rm out}(t)}\right)^2+
\left(\frac{y'}{w_{\rm out}(t)}\right)^2+
\left(\frac{z'}{w_{\rm out}(t)}\right)^2
\right]^{1/2}.
\]
The cocoon corresponds to the region
\[
R_{\rm in}\leq 1,
\]
whereas the swept-up shell is represented by
\[
R_{\rm in}>1 \quad \mathrm{and} \quad R_{\rm out}\leq 1.
\]

For each emitting component, we assume that the emissivity is spatially uniform within its corresponding volume. Thus, for a given spectral luminosity $L_\nu(t)$, the volume emissivity of a component $k$ is
\begin{equation}
    j_{\nu,k}(t) = \frac{L_{\nu,k}(t)}{V_k(t)},
\end{equation}
where $k=\{\mathrm{c},\mathrm{sh}\}$ denotes cocoon or shell, respectively. The projected specific intensity is then obtained by integrating the emissivity along the line of sight (los):
\begin{equation}
    I_\nu(x,y;t) =
    \frac{1}{4\pi}
    \int_{\rm los}
    \left[
    j_{\nu,\rm c}(t)\,\mathcal{H}_{\rm c}
    +
    j_{\nu,\rm sh}(t)\,\mathcal{H}_{\rm sh}
    \right]\,{\rm d}z,
    \label{eq:map_Inu}
\end{equation}
where $\mathcal{H}_{\rm c}$ and $\mathcal{H}_{\rm sh}$ are functions that are equal to one inside the cocoon and shell regions, respectively, and zero elsewhere. The factor $1/4\pi$ accounts for isotropic emission. The spatial coordinates are finally converted into angular coordinates using $\alpha=x/d$ and $\beta=y/d$, where $d$ is the distance to the source.

For the radio maps, we compute the specific intensity at a fixed observing frequency $\nu_0$. In this band we include only synchrotron radiation: the cocoon contribution is produced by primary electrons confined inside the cocoon, whereas the shell contribution includes synchrotron emission from both secondary leptons and primary electrons that have escaped from the cocoon. From the computed SEDs, we obtain the spectral luminosity as
\begin{equation}
    L_{\nu,k}(\nu_0,t)
    =
    \frac{\nu L_{\nu,k}(\nu_0,t)}{\nu_0}.
\end{equation}
The corresponding intrinsic radio map is then calculated from Eq. \eqref{eq:map_Inu}.

For the X-ray maps, we integrate the model luminosity over the soft X-ray band between $E_1=0.5$ keV and $E_2=2$ keV. The band luminosity of each component is
\begin{equation}
    L_{X,k}(t)
    =
    \int_{E_1}^{E_2}
    L_{E,k}(E,t)\,{\rm d}E .
\end{equation}
The shell X-ray luminosity includes thermal Bremsstrahlung from the shocked ISM, as well as non-thermal Bremsstrahlung and synchrotron emission from leptons in the shell. The cocoon X-ray luminosity includes synchrotron and inverse Compton emission from electrons confined within the cocoon. The corresponding band-integrated emissivity is
\begin{equation}
    j_{X,k}(t)=\frac{L_{X,k}(t)}{V_k(t)},
\end{equation}
and the projected soft X-ray intensity is
\begin{equation}
    I_X(x,y;t)
    =
    \frac{1}{4\pi}
    \int_{\rm los}
    \left[
    j_{X,\rm c}(t)\,\mathcal{H}_{\rm c}
    +
    j_{X,\rm sh}(t)\,\mathcal{H}_{\rm sh}
    \right]\,{\rm d}z .
    \label{eq:xray_map}
\end{equation}
We do not include interstellar absorption in the synthetic X-ray maps of this work. This effect can be incorporated later through an energy-dependent attenuation factor, $\exp[-\sigma(E)N_{\rm H}]$, when comparing the model with a particular source or line of sight.

\end{appendix}

\end{document}